\documentclass[aps,pra,twocolumn,superscriptaddress,showpacs,amsmath]{revtex4}
\usepackage{graphicx,hyperref}

\def\NAT@sort{1}
\newcommand{\ml}{\mathcal}

\newcommand{\bs}{\boldsymbol}

\newcommand{\bra}{\langle}
\newcommand{\ket}{\rangle}

\hypersetup{colorlinks=true,linkcolor=blue,citecolor=red,urlcolor=blue}

\begin{document}

\author{N. Kaplis}
\affiliation{Instituut-Lorentz for Theoretical Physics, Leiden University, Niels Bohrweg 2, Leiden 2333 CA, The Netherlands}

\author{F. Kr\"uger}
\affiliation{London Centre for Nanotechnology, University College London, Gordon St., London WC1H 0AH, United Kingdom}
\affiliation{ISIS Facility, Rutherford Appleton Laboratory, Chilton, Didcot, Oxon, OX11 0QX, United Kingdom}

\author{J. Zaanen}
\affiliation{Instituut-Lorentz for Theoretical Physics, Leiden University, Niels Bohrweg 2, Leiden 2333 CA, The Netherlands}

\title{Entanglement entropies and fermion signs of critical metals}

\begin{abstract}
The fermion sign problem is often viewed as a sheer inconvenience that plagues numerical studies of strongly interacting electron systems. Only recently, it has been suggested that fermion signs are fundamental for the universal behavior of critical metallic systems and crucially enhance their degree of quantum entanglement. In this work we explore potential connections between emergent scale invariance of fermion sign structures and scaling properties of bipartite entanglement entropies. Our analysis is based on a wavefunction ansatz  that incorporates collective, long-range backflow correlations into fermionic Slater determinants. Such wavefunctions mimic the collapse of a Fermi liquid at a quantum critical point. Their nodal surfaces -- a representation of the fermion sign structure in many-particle configurations space -- show fractal behavior up to a length scale $\xi$ that diverges at a critical backflow strength. We show that the Hausdorff dimension of the fractal nodal surface depends on $\xi$, the number of fermions and the exponent of the backflow. For the same wavefunctions we numerically calculate the second R\'enyi entanglement entropy $S_2$. Our results show a cross-over from volume scaling, $S_2\sim \ell^\theta$ ($\theta=2$ in $d=2$ dimensions), to the characteristic Fermi-liquid behavior $S_2\sim \ell\ln \ell$ on scales larger than $\xi$. We find that volume scaling of the entanglement entropy is a robust feature of critical backflow fermions, independent of the backflow exponent and hence the fractal dimension of the scale invariant sign structure. 
\end{abstract}

\maketitle

\section{Introduction}
\label{sec:introduction}
\begin{figure}[t!]
\begin{center}
\includegraphics[width=  0.35\linewidth]{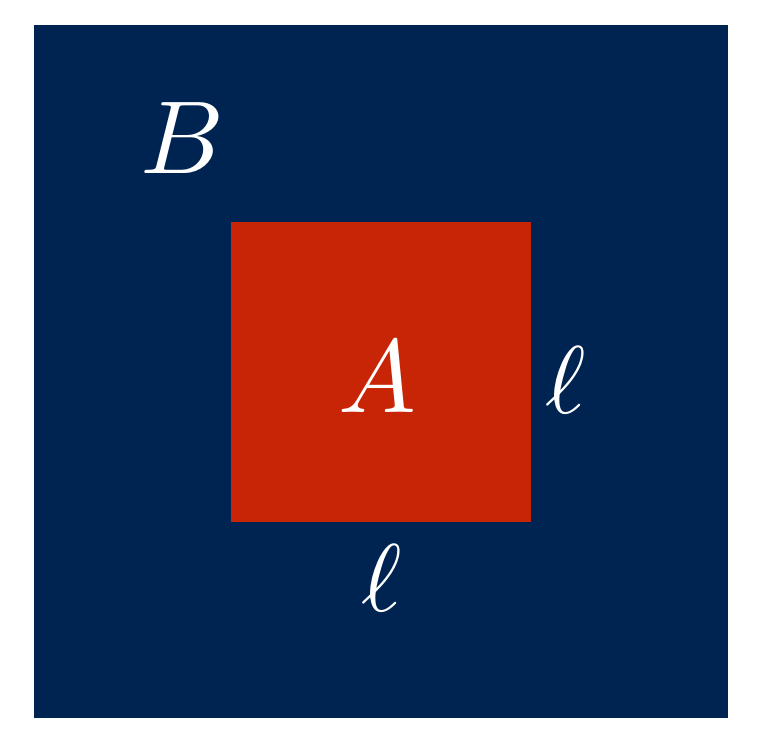}
\caption{(Color online) Two-dimensional system, divided into two sub-systems A,B. Sub-system B is traced out.}
\label{fig:entanglementArea}
\end{center}
\end{figure}

Bipartite entanglement entropies have attracted much attention as quantum information measures in the many body context (see Ref.~\cite{Laflorencie:2015eck} for a recent review). The idea is to divide the system into two spatial regions,  $A$ and $B$ (see Fig.\ref{fig:entanglementArea}), and to compute the reduced density matrix of subsystem $A$ by taking a partial trace on the full density matrix, $\hat{\rho}_A = \textrm{Tr}_B \hat{\rho}$. One can then compute the von Neumann entanglement entropy associated with the reduced density matrix by $S_{vN} = -\textrm{Tr} ( \hat{\rho}_A \ln \hat{\rho}_A )$, or alternatively, the $n$-th R\'enyi entropy by $S_n = \textrm{Tr} (\hat{\rho}_A^n)/(1-n)$. The Renyi and von Neumann entanglement entropies are  related to each other by the ``replica limit'', $S_{vN} = \lim_{n \rightarrow 1}  S_n$ and they constitute measures of how the states in $A$ are entangled with those in $B$.

In gapped systems, the entanglement entropy scales with the area of the interface between the sub-systems, $S_{vN} \sim S_n \sim \ell^{d-1}$, simply because correlations are short ranged. Perhaps surprisingly, this \emph{area law} turns out to be generic to bosonic ground states in dimensions $d\ge 2$. Its robustness has first been demonstrated in the context of free bosonic field theories \cite{Bombelli+86,Srednicki93,Callan+94}. More recently, it has been shown that Goldstone modes in ordered systems with broken continuous symmetry \cite{Kallin+11}, topological order in gapped systems \cite{Furukawa+07}, and the scale invariance at bosonic quantum-critical points \cite{Metlitski+09} only give rise to sub-leading, additive corrections to the area law. Across quantum phase transitions, cusp singularities are found in the pre-factor of the area law \cite{Helmes+14,Frerot+16}. Hence, the area law does not a priori carry information regarding a given phase of matter \cite{Laflorencie:2015eck,Chandran:2013zqa}. 

In sharp contrast to the celebrated area-law scaling in the quantum ground states, finite energy density eigenstates typically satisfy a \emph{volume law} scaling, $S_{vN} \sim S_n \sim \ell^d$ \cite{Page93,Foong+94,Sanchez95,Sen96}. Recently Grover and Fisher \cite{Grover:2013tia,Grover:2014yea} presented considerations that appear to shed light on the origin of the gross differences between (near) ground states and highly excited states, considering states of the form $| \Phi \rangle = \sum_i A_i |\textrm{config}, i \rangle$, where $|\textrm{config}, i \rangle$ are the configurations of the $N$-particle Hilbert space in a certain local basis. They show that various generic wave functions with all $A_i \ge 0$ cannot exceed an area-law  entanglement \cite{Grover:2014yea}. This is consistent with the area-law bound for bosonic ground states: as first realized by Feynman \cite{Feynman72,Wu09}, many-body ground-state wavefunctions of bosons in the coordinate representation are node-less and it is therefore possible to find a local basis in which all the amplitudes $A_i$ can be chosen to be positive definite. In an earlier paper \cite{Grover:2013tia}, Grover and Fisher consider states such that all $A_i$ have the same absolute value but {\em random signs}, with the effect that the entanglement entropy acquires a {\em volume} scaling. Such states should be representative for highly excited energy eigenstates, given that these will form a dense continuum while all states have to be orthogonal to each other which can only be accomplished when sign changes are maximally dense. These findings suggests that the scaling of the entanglement entropies is not determined by entanglement in a general sense but instead by the {\em sign structure} carried by the entangled state.

It is perhaps a universal affair that the sign structures of highly excited states are of such complexity that they can be regarded as random for all practical purpose. However, upon descending to low energy there might be room for more structure. This becomes especially relevant dealing with the plethora of problems that are characterized by the (fermion) sign problem. The vacuum states of interacting fermions at a finite density or generic quantum spin problems are characterized by sign changes that cannot be transformed away.  Systems suffering from such sign problems do no longer map on probabilistic systems and these are claimed to be in general non-computable: the computation of the vacuum is a NP hard problem \cite{Troyer:2004ge}. This fundamental fact appears to be overlooked in attempts to prove that in full generality the ground states associated with local Hamiltonians would be characterized by an area law for the entanglement entropy. 

The only genuine sign-full vacuum that is under complete control is the ground state of the Fermi gas and its perturbative extension, the Fermi liquid. This is characterized by the anti-symmetrization procedure explained in the quantum mechanics textbooks. This amounts to an irreducible long range entanglement in a real space representation only involving the exchange signs which has proven to be very difficult to encode in for instance a tensor network, given its non-local nature. It is a famous result that its entanglement entropy scales like area-log-area, $S_{vN} \sim S_n \sim \ell^{d-1}\ln\ell$ \cite{PhysRevLett.96.010404,PhysRevLett.96.100503,PhysRevB.74.073103,PhysRevA.74.022329,PhysRevB.87.081108}, i.e. longer ranged than a typical bosonic system \cite{bose_metal}. 

It is likely that non-Fermi liquid states such as quantum critical metals are characterized by sign structures that are radically different from those of conventional metals, giving rise to a very different, universal form of entanglement scaling. This intuition is inspired by the strange metallic states that are predicted by the holographic duality \cite{zaanen2015holographic}, known as the AdS/CFT correspondence. These states appear to correspond generically to {\em quantum critical phases} characterized by an emergent scale invariance which does not require fine tuning to critical points. It has been demonstrated \cite{PhysRevB.85.035121} that such holographic strange metals can exhibit an anomalous entanglement-entropy scaling $S_{vN} \sim \ell^{\theta}$ with an exponent that can take any value $d - 1 < \theta \le d$ and that is equal to the hyperscaling violation exponent \cite{PhysRevLett.56.416}. While there is abundant evidence that the resulting vacua are ``sign-full'' non-Fermi liquids, their sign-structures have not been investigated yet.
 
In the paper, we explore the connection between non-trivial sign structures and the scaling of bi-partite entanglement entropies in the context of fermionic hydrodynamical backflow wave functions. It is not known whether these are eigenstates of any realistic Hamiltonian but they have quite a history as a device to wire in a richer sign structure in numerical quantum Monte-Carlo computations 
\cite{PhysRevLett.47.807,PhysRevB.48.12037,PhysRevB.58.6800,PhysRevE.68.046707,PhysRevB.78.041101}. The highlight is that, in a certain parameter regime, backflow wavefunctions are characterized by a {\em fractal distribution} of zeros in configuration space \cite{2008PhRvB..78c5104K}.  This backflow system was therefore introduced as a model that might shed light on the nature of quantum critical non-Fermi liquid states \cite{2008PhRvB..78c5104K}, revolving around the conjecture that a fractal nodal surface is a necessary condition for the emergence of scale invariance in any system characterized by ``irreducible'' signs in the ground state.

The outline of this paper is as follows. In Sec.~\ref{sec:backflow}, we review the fermionic backflow wave-function ansatz and introduce the surface of zeros of the wave function in configuration space as a geometrical measure of the sign structure. We illustrate the emergent scale invariance in the nodal structure and show that by generalizing the wave-function ansatz used in previous work \cite{2008PhRvB..78c5104K}, it is possible to vary the fractal dimension of the nodal surface 
over a significant range. In Sec.~\ref{sec:results} we numerically calculate the R\'enyi entanglement entropy $S_2$ for the states described by the generalized backflow wavefunctions in two dimensions. Our main finding is that irrespective
of the fractal dimension of the nodal surface of the critical backflow state, the entanglement entropy follows a volume law. In Sec.~\ref{sec:disc}, we summarize our results and discuss their implications.

\section{Sign structures of backflow fermions}
\label{sec:backflow}
Fermion signs are expected \cite{2008PhRvB..78c5104K} to play a crucial role for the universal behavior at quantum critical points. This insight is based on the constrained world-line path integral reformulation \cite{Ceperley91} of the sign-full fermionic path integral, which is used in quantum Monte-Carlo simulations to treat the fermion signs in a more manageable way \cite{PhysRevLett.69.331,PhysRevLett.73.2145,PhysRevLett.76.1240}. In this language the nodes of the many particle density matrix impose hard-core constraints on the dynamics of effectively bosonic world lines. This \emph{nodal hypersurface} can be viewed as a geometrical representation of the fermion sign structure in many-particle configuration space.

The Fermi energy $E_F$ is encoded in the constraint structure \cite{2008PhRvB..78c5104K}: the average nodal pocket size gives rise to an average collision time $\tau_c\sim 1/E_F$ of the worldlines with the constraint structure. The observation of Planckian dissipation in the quantum critical region of high-$T_c$ cuprates \cite{Marel+03}, as well as the discontinuous Fermi-surface reconstruction \cite{2004Natur.432..881P} and the quasi-particle divergence \cite{Custers+03} seen in heavy-fermion intermetallics clearly show that at the quantum critical point the metallic system loses its knowledge of the Fermi degeneracy scale. From the above considerations based on the constrained path integral it is therefore clear that the fermion sign structure has to become scale invariant.

It has been demonstrated \cite{2008PhRvB..78c5104K} that collective, long-range backflow correlations built into fermionic wavefunctions can lead to emergent scale invariance in the nodal structure that goes hand in hand with a disappearance of the discontinuity in the single-particle momentum distribution $n(\bs{k})$. Fermionic backflow wavefunctions therefore provide a simple tool to study the collapse of a Fermi liquid at a critical point and to investigate how the Hausdorff dimension of the fractal nodal structure enters scaling relations. In this section we briefly review the properties of the fermionic backflow gas. We will then generalize the wavefunction ansatz used in previous work \cite{2008PhRvB..78c5104K} and demonstrate that the fractal dimension can be tuned over a significant range, e.g. by changing the exponent of the long-range backflow correlations.

The idea to incorporate hydrodynamical backflow effects in quantum-mechanical wave functions dates back to Feynman and Cohen \cite{PhysRev.102.1189}. They argued that the `roton' in Helium-4 is like a single mobile atom which is dressed by collective motions in the liquid. Helium is a nearly incompressible fluid and the density in the neighborhood of the moving particle is barely altered. As a consequence, there has to be a backflow of other particles conserving the total current and leading to an enhancement of the effective mass of this quasiparticle. This can be described quantum mechanically by taking
wave functions $\exp(i\bs{k}\tilde{\bs{r}}_i)$ with collective quasiparticle coordinates
\begin{equation}
\tilde{\bs{r}}_i=\bs{r}_i+\sum_{j(\neq i)}\eta(|\bs{r}_i-\bs{r}_j|)(\bs{r}_i-\bs{r}_j),
\label{eq:coll}
\end{equation}
where the $\bs{r}_i$ are the coordinates of the bare particles and $\eta(r)$ is a smoothly varying function which falls off like $\sim r^{-3}$ on large distances, exactly as the dipolar backflow in a classical, incompressible fluid. Much later, it was found out that wave functions of the form $\Psi_\textrm{bf}= J(\bs{r}_1,\ldots,\bs{r}_N) \det\left(e^{i\bs{k}_i\tilde{\bs{r}}_j}  \right)$, with long-range back-flow correlations built into the Slater determinant, yield excellent variational Monte-Carlo energies for bulk liquid $^3$He \cite{PhysRevLett.47.807}, electron jellium \cite{PhysRevB.48.12037,PhysRevB.58.6800}, metallic Hydrogen \cite{PhysRevE.68.046707}, and the two-dimensional Hubbard model \cite{PhysRevB.78.041101}.
\begin{figure}[t!]
\begin{center}
\includegraphics[width= 0.95\linewidth]{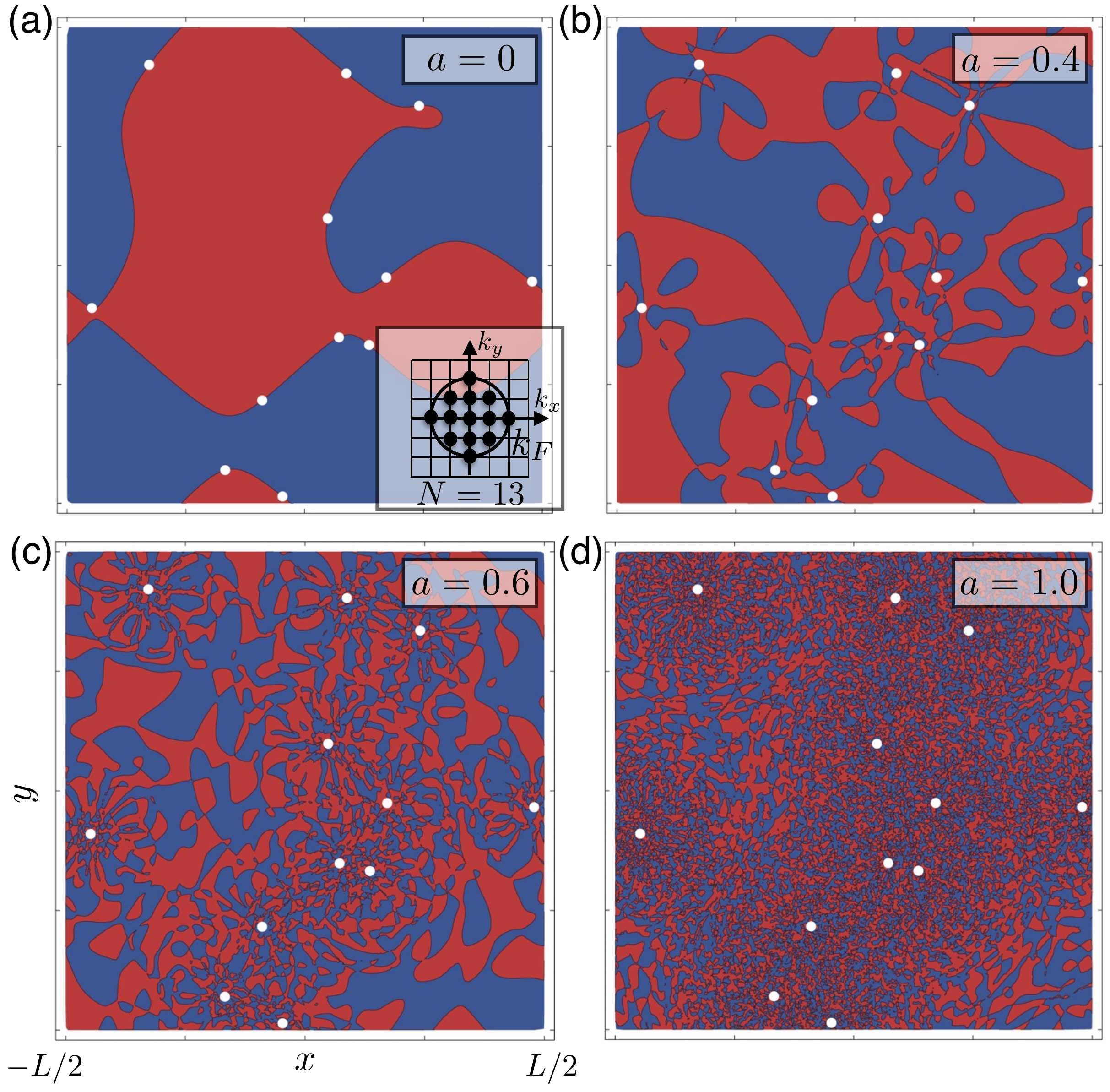}
\caption{(Color online) Two-dimensional cuts of the nodal surface of fermionic backflow wavefunctions for $k_F=2$ ($N=13$),  backflow exponent $\beta=3$, and increasing backflow strength $a$. The white dots show the fixed positions of $N-1$ particles. The nodal surface seen by the remaining particle is given by the interface between blue and red areas, corresponding to the two different signs of the wavefunction.}
\label{fig:nodalSurfaces}
\end{center}
\end{figure}
Note that the Jastrow factors $J$ are positive definite and symmetric under particle exchange. Hence, the fermion sign structure is completely determined by the determinant factor and will crucially depend on the collective backflow.  Following Ref.~\cite{2008PhRvB..78c5104K}, we ignore the Jastrow factor (as well as the spin dependence) and consider backflow wave functions of the form
\begin{equation}
\Psi(\bs{r}_1,\ldots,\bs{r}_N) = \ml{N} \det\left( e^{\imath \bs{k}_i\cdot \tilde{\bs{r}}_j}\right),
\label{eq:waveFunction}
\end{equation}
where  $\ml{N}$ denotes a normalization factor and the collective coordinates are defined in Eq.~(\ref{eq:coll}). We will use a generalized backflow function
\begin{equation}
\eta(r) = a^\beta/\left(r^\beta + r_0^\beta\right),
\label{eq:backflow}
\end{equation}
where $a$ determines the strength of the backflow and $r_0$ is a short-distance cut-off. The backflow exponent $\beta=3$ encodes for the literal hydrodynamical, dipolar backflow. We will also use different values of $\beta$ and show that this amounts to a flexible way to change the fractal dimension of the critical nodal surface. This represents the key result of this section. On very small distances, the fractal behavior is cut-off by $r_0$. All our results are independent of the particular choice of $r_0$, as long as it is chosen to be sufficiently small compared to the other length scales in the system. 

\begin{figure}[t]
\begin{center}
\includegraphics[width= 0.95\linewidth]{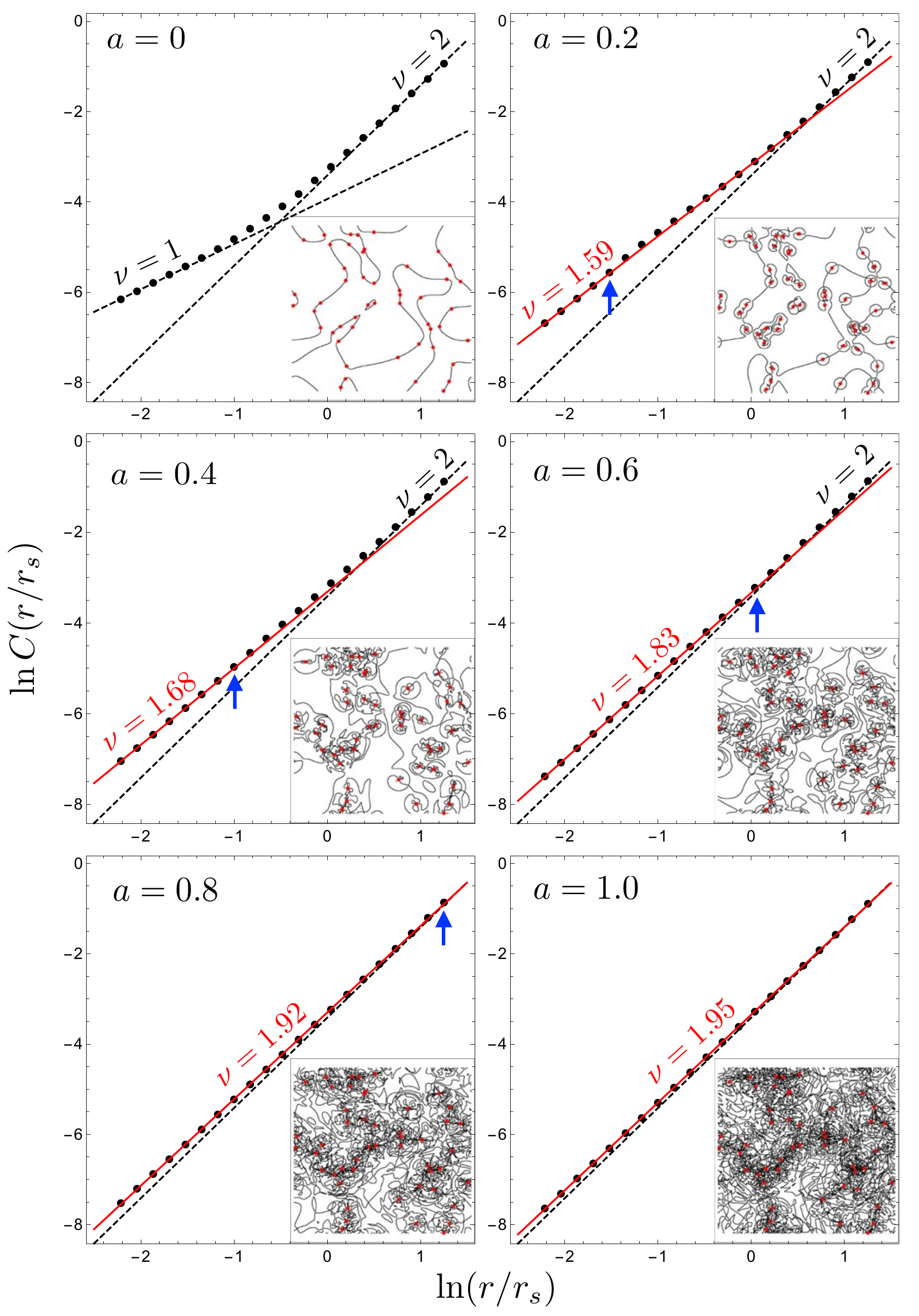}
\caption{(Color online) Correlation integrals for $N=49$, $\beta=3$, and different values of the backflow strength $a$. Corresponding nodal-surface cuts are shown as insets. Both the range $\xi$ of 
fractal behavior (indicated by blue arrows) and the Hausdorff dimension $\nu$ increase with $a$. For $a=1.0$, $\xi$ exceeds the system size.}
\label{fig:correlationIntegrals}
\end{center}
\end{figure}

The wavefunctions (\ref{eq:waveFunction}) are eigenstates of a free-particle Hamiltonian in terms of the collective coordinates, $\mathcal{H}=-\frac{\hbar^2}{2m}\sum_i \partial^2/\partial \tilde{\bs{r}_i}^2$. In the following we will consider the ground state of this backflow gas for a two-dimensional square system of size $L\times L$ with periodic boundary conditions. This corresponds to a set of momenta $\bs{k}_i$ on a two dimensional grid with spacing $\Delta k=2\pi/L$ and $|\bs{k}_i|\le k_F$,  e.g. for $k_F=2$ (in units of $\Delta k$) we obtain $N=13$ particles with momenta shown in the inset of Fig.~\ref{fig:nodalSurfaces}(a).

The nodal hypersurface is determined by the zeros of the wavefunction, $\Psi(\bs{r}_1,\ldots,\bs{r}_N)=0$. The geometry and topology of this object characterize the sign structure of the fermionic state. In Fig.~\ref{fig:nodalSurfaces} we show two-dimensional cuts, which are obtained by fixing $N-1$ particles at random positions. For the non-interacting Fermi gas the nodal surface seen by the remaining particle smoothly connects the $N-1$ fixed particles, forming pockets with a typical size of the order of the inter-particle spacing [Fig.~\ref{fig:nodalSurfaces}(a)]. Note that for any fermionic wavefunction the $N-1$ particles are necessarily located on the nodal-surface cut because of Pauli's exclusion principle. Figs.~\ref{fig:nodalSurfaces}(b)-(d) show the evolution of the nodal surface as a function of the backflow strength $a$ for  $\beta=3$.

In order to quantify the changes of the nodal surface and to demonstrate that backflow indeed leads to fractal behavior, we numerically calculate the correlation integral $C(r)$ which counts the number of pairs of nodal-surface points with separation less than $r$. For a fractal object, $C(r)\sim r^\nu$ with $\nu=d_H$ the Hausdorff or fractal dimension \cite{2008PhRvB..78c5104K}. The resulting correlation integrals for $N=49$, $\beta=3$ and different values of the backflow strength $a$ are shown in Fig.~\ref{fig:correlationIntegrals}. Without backflow, we find $\nu=1$ on scales smaller than the inter-particle spacing $r_s$, consistent with smooth one-dimensional nodal lines. Near $r_s$ there exists a crossover to $\nu=2$, reflecting that the nodal structure looks two-dimensional on scales larger than the average spacing between nodal lines. For $a>0$ the correlation integrals show fractal behavior from the small-distance cut-off $r_0$ up to a `correlation-length' scale $\xi$ which increases with $a$ and seems to diverge at a critical value $a_c\approx 0.9-1.0$. At this value, the back-flow strength $a$ is of the order of the inter-particle spacing. Unfortunately, it is not possible to extract accurate values of $\xi$ and $a_c$ since the crossover is relatively broad for the system sizes we can study and since changes of $\nu$ become quite small near $a_c$.
\begin{figure}[t!]
\begin{center}
\includegraphics[width=0.7 \linewidth]{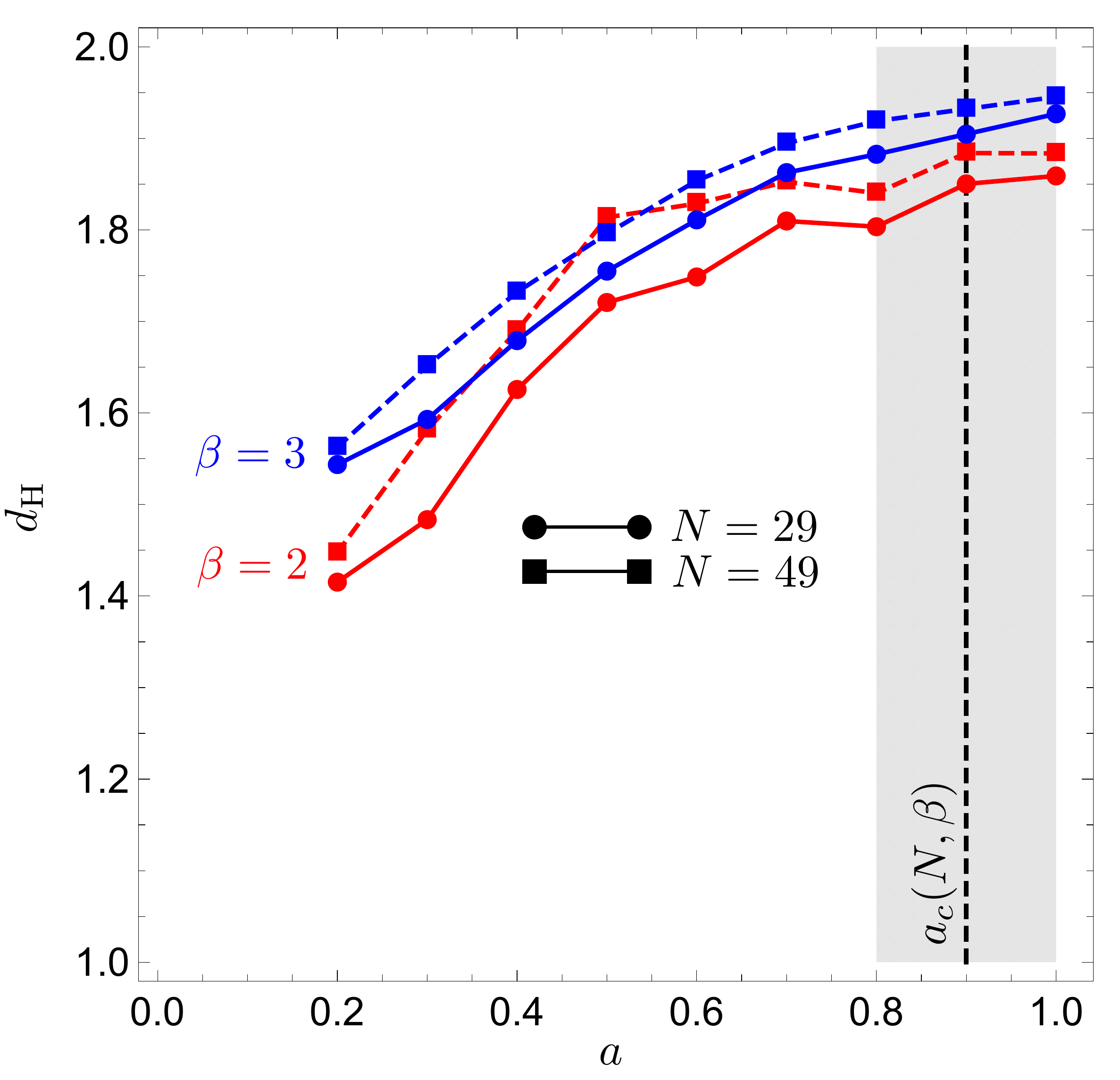}
\caption{(Color online) Fractal dimension $d_H=\nu$ of the nodal surface as a function of the backflow strength $a$ for backflow exponents $\beta=2$ (red) and $\beta=3$ (blue) and different particle numbers $N$.}
\label{fig:hausdorffDimensions}
\end{center}
\end{figure}
Interestingly, the fractal dimension $\nu$ is found to be non-universal and to increase with $a$ over quite a big range. This behavior is summarized in Fig.~\ref{fig:hausdorffDimensions} for different particle numbers $N$ and back-flow exponents $\beta$. As expected, the Hausdorff dimension depends on $\beta$. It also changes noticeable between $N=29$ and $N=49$, indicative of a relatively strong finite-size dependence.

In our earlier work \cite{2008PhRvB..78c5104K} we already observed that the Hausdorff dimension $d_H$ of the nodal surface does depend on $a$, but only considered the canonical value $\beta = 3$ for the backflow exponent.  Although the identification with the physical backflow notion is no longer possible, one can just exploit the freedom in the parametrization  (\ref{eq:backflow}) as defining a family of wave-function {\em ansatzes}. 

As it turns out, for a ``longer ranged" envelope function ($\beta < 3)$, $d_H$ decreases while it increases for $\beta >3$. A typical example is shown in Fig.~\ref{fig:hausdorffDimensions}: one infers that relative to the $\beta =3$ results, the Hausdorff dimensions are significantly smaller for $\beta =2$. We will exploit the differences of these two cases in our study of the R\'enyi entropies. However, we have explored a larger family of envelope functions, finding that the Hausdorff dimension can in principle be tuned to have any value $1<d_H<2$.

\section{The second Renyi entropy of backflow fermions}
\label{sec:results}

Let us now turn to the results for the second R\'enyi entropy $S_2$ for the fermionic backflow wavefunctions (\ref{eq:waveFunction}), calculated numerically by using the Monte-Carlo algorithm \cite{PhysRevLett.107.067202,Hastings+10} outlined in Sec.~\ref{sec:Renyi}. The wave functions are defined on a square system of side-length $L$ with periodic boundary conditions. We choose the sub-system $A$ to be an $\ell\times\ell$ square in the center of the system (see Fig.~\ref{fig:entanglementArea}). We will compute $S_2(\ell/L)$ for different particle numbers $N$, backflow strengths $a$, and for different values of the backflow exponent $\beta$. As we have shown in Sec.~\ref{sec:backflow}, $\beta$ serves as a `knob' to change the Hausdorff dimension of the fractal nodal surface. 

Before turning to the effects of backflow, let us first benchmark our code by calculating the entanglement entropy for free fermions ($a=0$). We expect that $S_2\sim \ell \ln \ell$, at least in the regime $z:=\ell/L\ll 1$. For larger $z$, finite-size effects will start to become important. Since by construction the ground-state is non-degenerate, $S_2$ vanishes as $z\to1$. At around $z=1/\sqrt{2}$ where the area of sub-system $A$ is half of the total area of the system the entanglement entropy $S_2(z)$ has a maximum. In addition, commensuration effects between the partitioning and the periodic boundary conditions are know \cite{Calabrese+12} to cause small ripples or oscillations in $S_2(z)$ that are most pronounced near the maximum of $S_2$. To minimize such finite-size effects, we will investigate entanglement entropies only for values $z<1/2$.  
\begin{figure}[h]
\includegraphics[width= \linewidth]{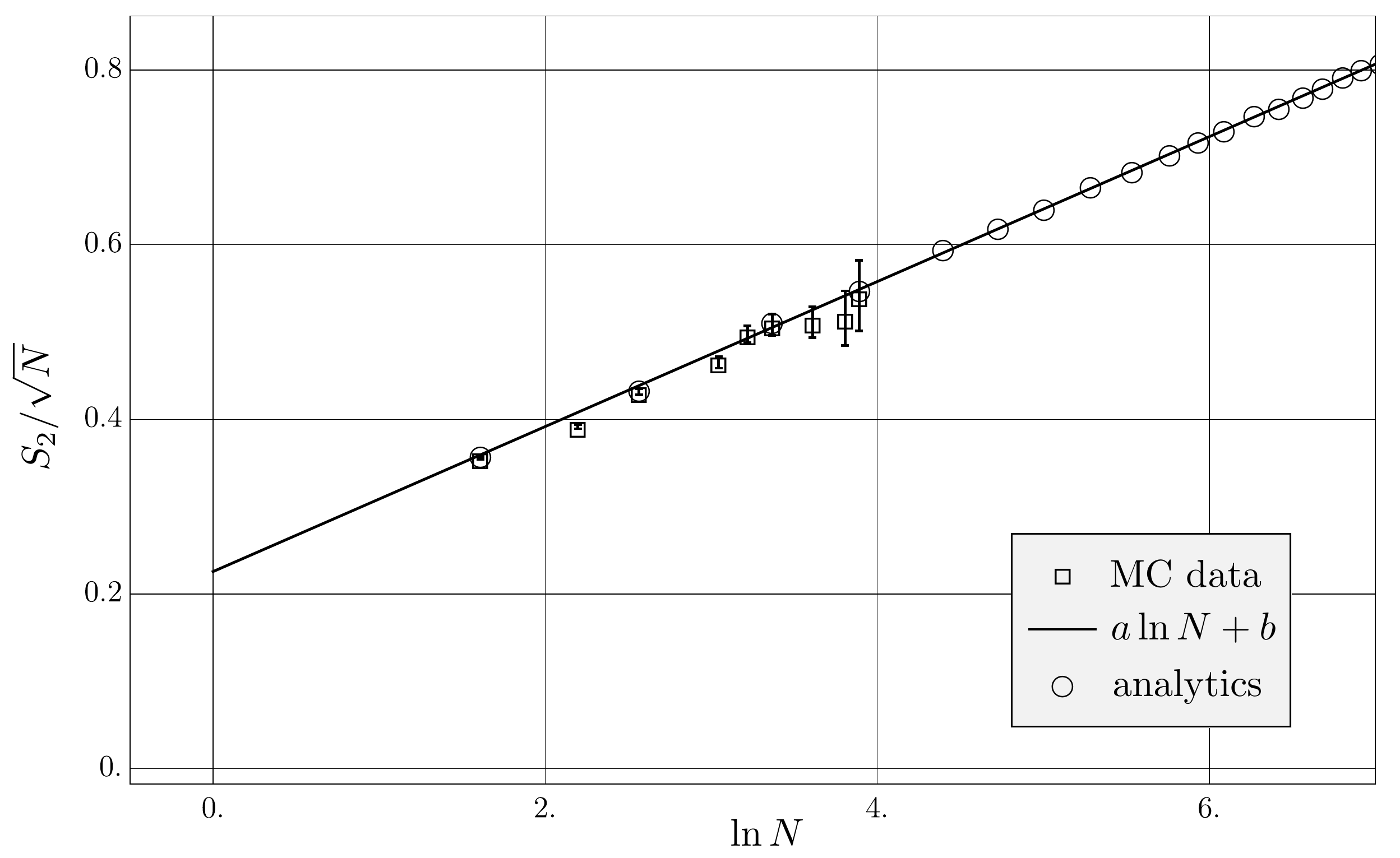}
\caption{Scaling of the second R\'enyi entropy vs. the number of particles (or equivalently the size of the system), for the non-interacting Fermi gas ($a=0$) and $z=0.3$. For comparison, we also show the analytic results obtained from 
the eigenvalues of the overlap matrix of single-particle states on the sub-volume.}
\label{fig:finiteNfree}
\end{figure}
We can look for the typical area-log-area Fermi-liquid scaling either as a function of $z$ or as a function of the particle number $N$. For fixed density $\rho=N/L^2$, one expects $S_2\sim \sqrt{N}\ln N$. However, it is known that for free fermions,
the entanglement entropy contains a seizable sub-leading area-law contribution $\sim \sqrt{N}$. In Fig.~\ref{fig:finiteNfree} we demonstrate that our MC data indeed follow the expected dependence $S_2/\sqrt{N} = a \ln N+b$ with coefficients 
$a$, $b$ that depend on the particular value of $z$.

For comparison, we have also calculated the entanglement entropies $S_2$ of the free fermion groundstate analytically. This requires the computation of the eigenvalues $\lambda_1,\ldots,\lambda_N$ of the overlap matrix 
$A_{ij} = \int_A \phi_i^*(\bs{r}) \phi_j(\bs{r})$ between the single-particle states $\phi_i(\bs{r})=L^{-1} e^{i\bs{k}_i\cdot\bs{r}}$ on the sub-system $A$. The entanglement entropy is obtained as $S_2=-\sum_i \ln \left[ \lambda_i^2+(1-\lambda_i)^2 \right]$ \cite{Calabrese+12}. As illustrated in Fig.~\ref{fig:finiteNfree}, we find excellent quantitative agreement between the analytic results and out MC data.

We now look in detail at the effects of backflow, starting with a backflow exponent of $\beta=3$ and varying the backflow strength $a$. As demonstrated in Sec.~\ref{sec:backflow}, backflow leads to fractal behavior in the nodal structure up to a length scale $\xi$. This length scale rapidly increases with $a$ and becomes of the order of the system size at $a_c\approx 0.9$, indicative of a diverging correlation length. One would expect to see a change in the entanglement-entropy scaling from critical, non-Fermi liquid behavior on scales $\ell<\xi$ to conventional Fermi-liquid behavior for $\ell>\xi$. As we will see below, this is indeed the case. 

\begin{figure}[t]
\centering
\includegraphics[width= \linewidth]{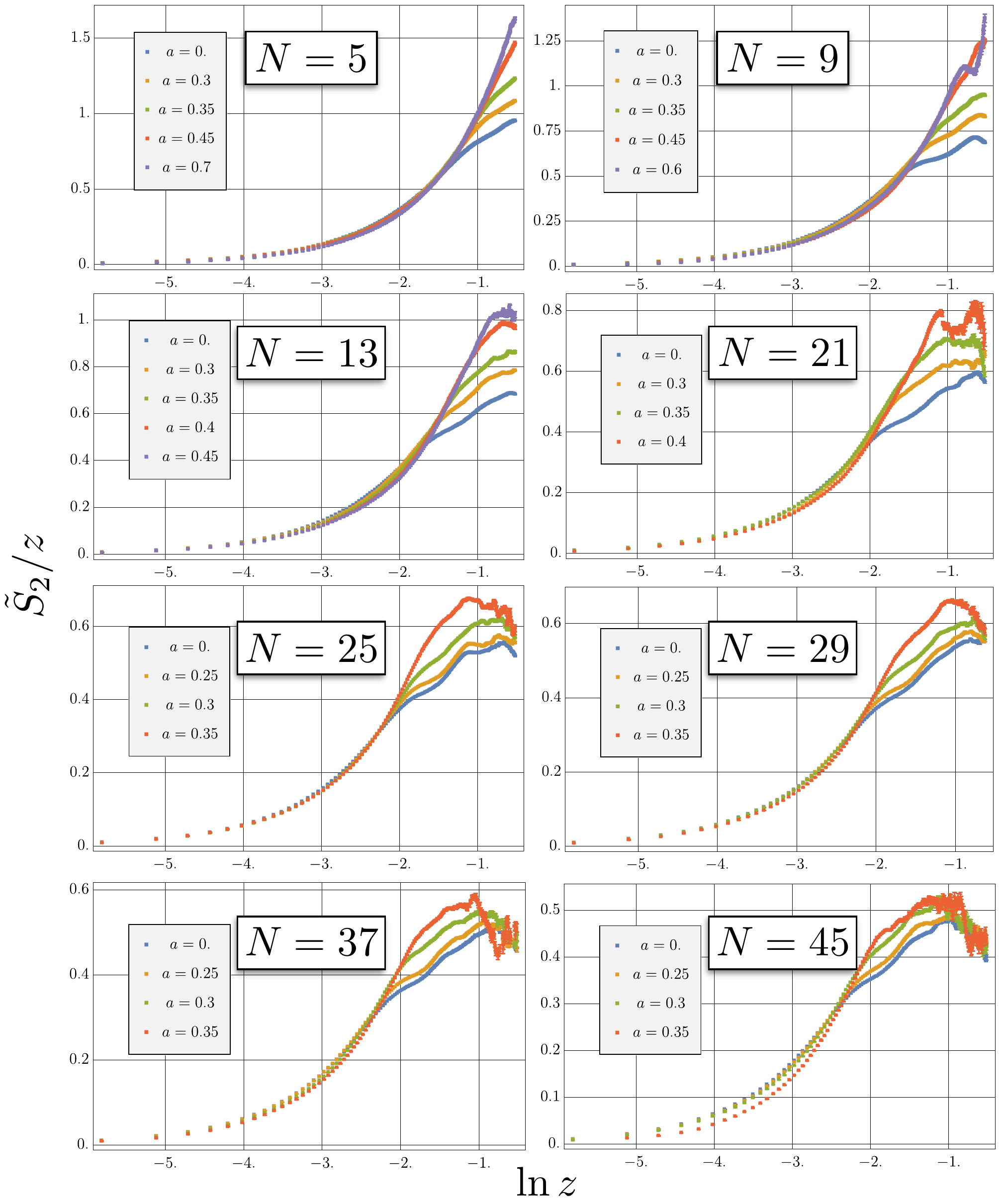}
\caption{(Colour on-line) Second R\'enyi entropy for the fermionic backflow system ($\beta=3$), for increasing number of particles $N$, as a function of $z$. Increasing values of $a$ are overlaid. We have defined $\tilde{S}_2=S_2/(\sqrt{N}\ln N)$.}
\label{fig:S2vsa}
\end{figure}
The results for the second R\'enyi entropy $S_2$ of the backflow system with given $N$ and $a$ as a function of $z=\ell/L$, are displayed in Fig.~\ref{fig:S2vsa}. A crossover in the entanglement-entropy scaling manifests itself as inflection point in $S_2(z)$. The location of this inflection point shifts to larger values of $z$ as the backflow strength is increased, until our view is obscured by finite-size effects. Note that the inflection point in the case of the free Fermi gas is entirely due to finite-size effects and not indicative of a crossover in the entanglement scaling. As we will demonstrate later, for $a>0$ the entanglement entropy follows a \emph{volume} law, $S_2\sim \ell^2$, on scales smaller than the length scale marked by the inflection point.

\begin{figure}[t]
\centering
\includegraphics[width=0.9\linewidth]{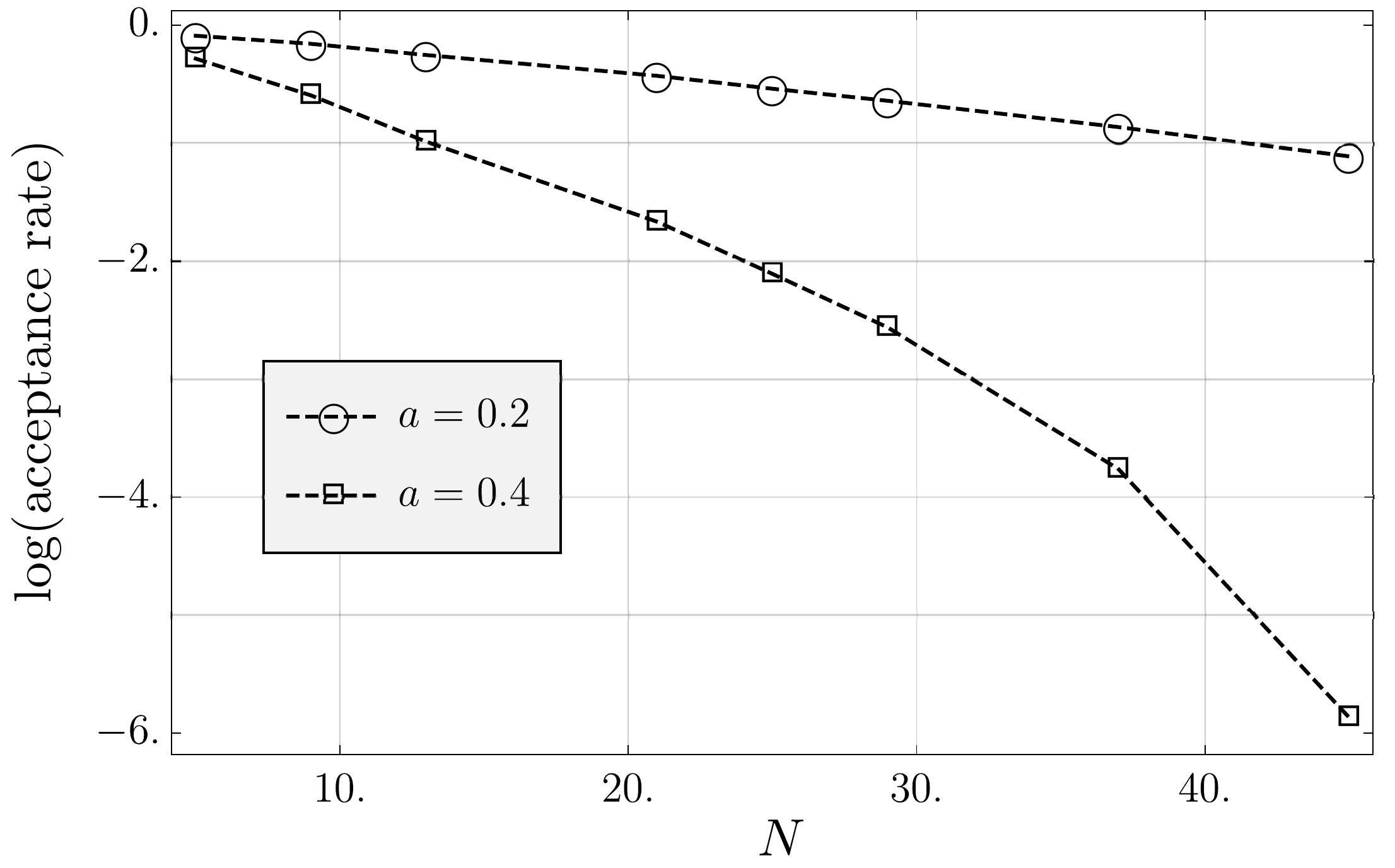}
\caption{The average acceptance rate for the Metropolis moves, as a function of the number of particles, for two values of the backflow strength, $a=0.2$ and $a=0.4$.}
\label{fig:acceptanceRate}
\end{figure}

At this point it should be noted that we are only able to calculate the entanglement entropies for backflow systems of up to $N=45$ particles. Already for $N=49$ the noise increases significantly. For even larger systems the calculation becomes impossible. This restriction persisted even though our calculations were parallelized on the Dutch GRID, which allowed us for almost unlimited CPU resources. What we observed was that the calculations become increasingly time-consuming as the acceptance rate, for each move proposed in the Metropolis algorithm, gets exponentially suppressed as the number of particles increases. This is amplified for more strongly interacting systems, i.e. for larger values of the backflow strength $a$ (see Fig.~\ref{fig:acceptanceRate}). This is reminiscent of the critical slowing down observed in other Quantum Monte Carlo simulations upon approaching a critical state.

Given the restricted number of particles available one naturally worries about the reliability of our results against finite-size artifacts. After all, in bosonic systems one usually spans several orders of magnitude of system sizes  in order to establish the thermodynamic limit.  To address this issue and to demonstrate the amazing resilience of the entanglement-entropy scaling of the fermionic backflow system to finite-size effects we study the dependence of our results on $N$. For this purpose we fix the density $\rho=N/L^2$ of the system as well as the size $\ell$  of the sub-system and vary the total size of the system. In this way we can isolate the effect of the size of the total system and examine its effects on the entropy. More importantly it means that the comparison of $S_2$ for different $N$, done this way, is meaningful. 
\begin{figure}[t]
\includegraphics[width=0.9\linewidth]{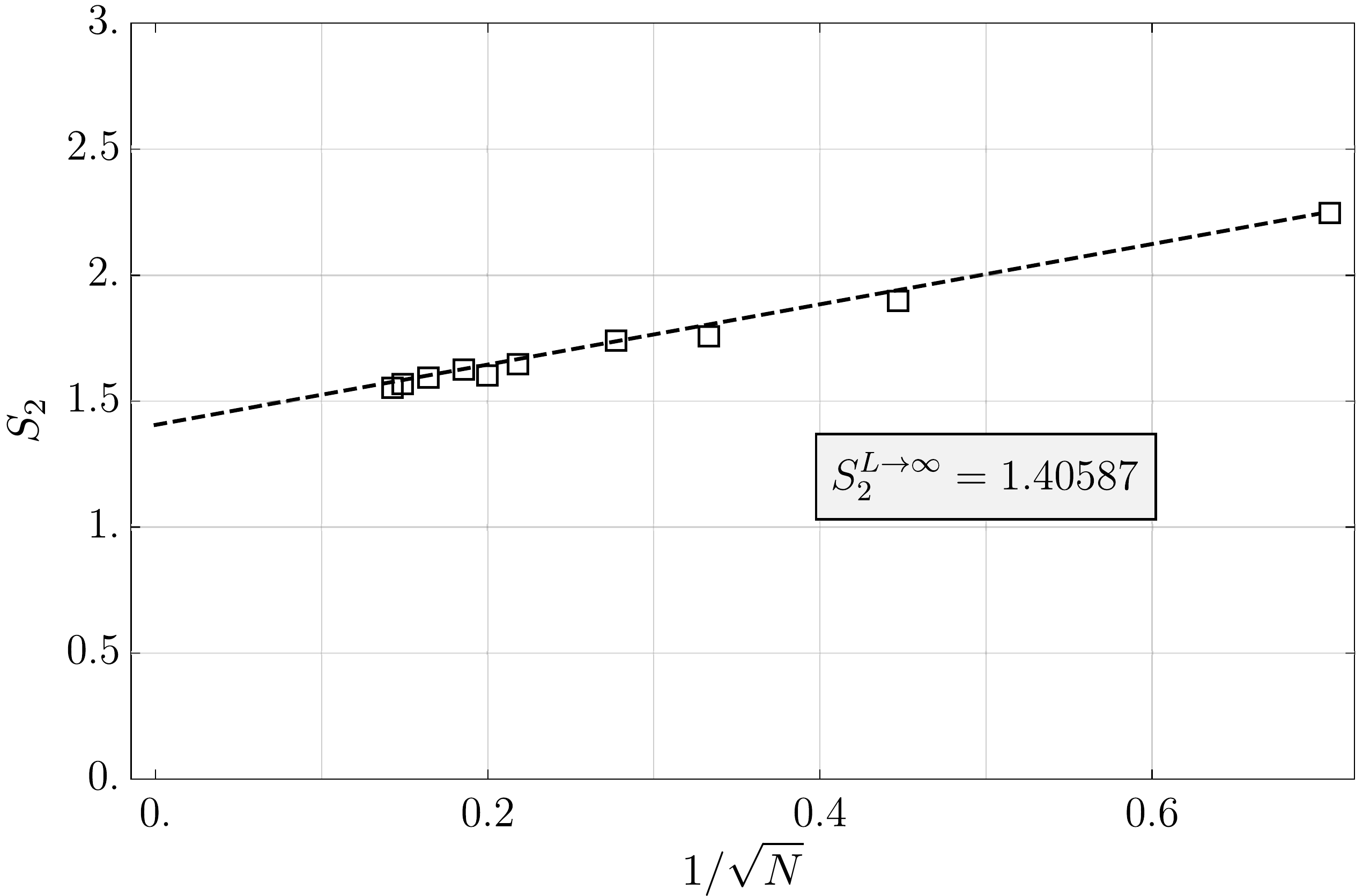}
\caption{The dependence of $S_2$ on the size of the system, keeping the density and the sub-system size constant. The entanglement entropies calculated for systems with a small number of particles are surprisingly close 
to the value for an infinite system.}
\label{fig:finiteN}
\end{figure}

It should be noted here that as the number of particles or the system size increases, the critical backflow strength $a_c$ decreases, resulting into noisier $S_2(z)$ curves. The fortunate caveat though is that as $N$ grows while $\rho$ and $\ell$ are kept fixed, $z$ becomes smaller. As it can be seen in Fig.~\ref{fig:S2vsa}, the lower values of $z$ remain relatively reliable even for larger $N$. This is what gives us confidence in the scaling of $S_2$ vs. $N$, presented in Fig.~\ref{fig:finiteN}. It is quite remarkable that the convergence rate to the thermodynamic limit is very fast and that even in very small systems, e.g. for $N=5$ particles, one seems to be able to capture the essential physical behavior. Unfortunately, even for only $N=2$ fermions with backflow correlations, an analytic calculation of $S_2$ appears to be impossible.

We can now proceed to our ultimate goal, which is to compute the scaling behavior $S_2\sim \ell^\theta$ of the R\'enyi entanglement entropy of a critical backflow gas, $a\to a_c$. This can be done by repeating the above scaling analysis for different values of $\ell$. We find very  good evidence that for $a<a_c$ and sufficiently small $z$, the R\'enyi entropy always follows a \emph{volume} law, $\theta=2$, irrespective of the value of $a$. This is in stark contrast to the behavior of the fractal dimension $d_H$ of the nodal surface which increases with $a$ over a relatively large range (see Fig.~\ref{fig:hausdorffDimensions}). 

\begin{figure}[t]
\includegraphics[width=0.95\linewidth]{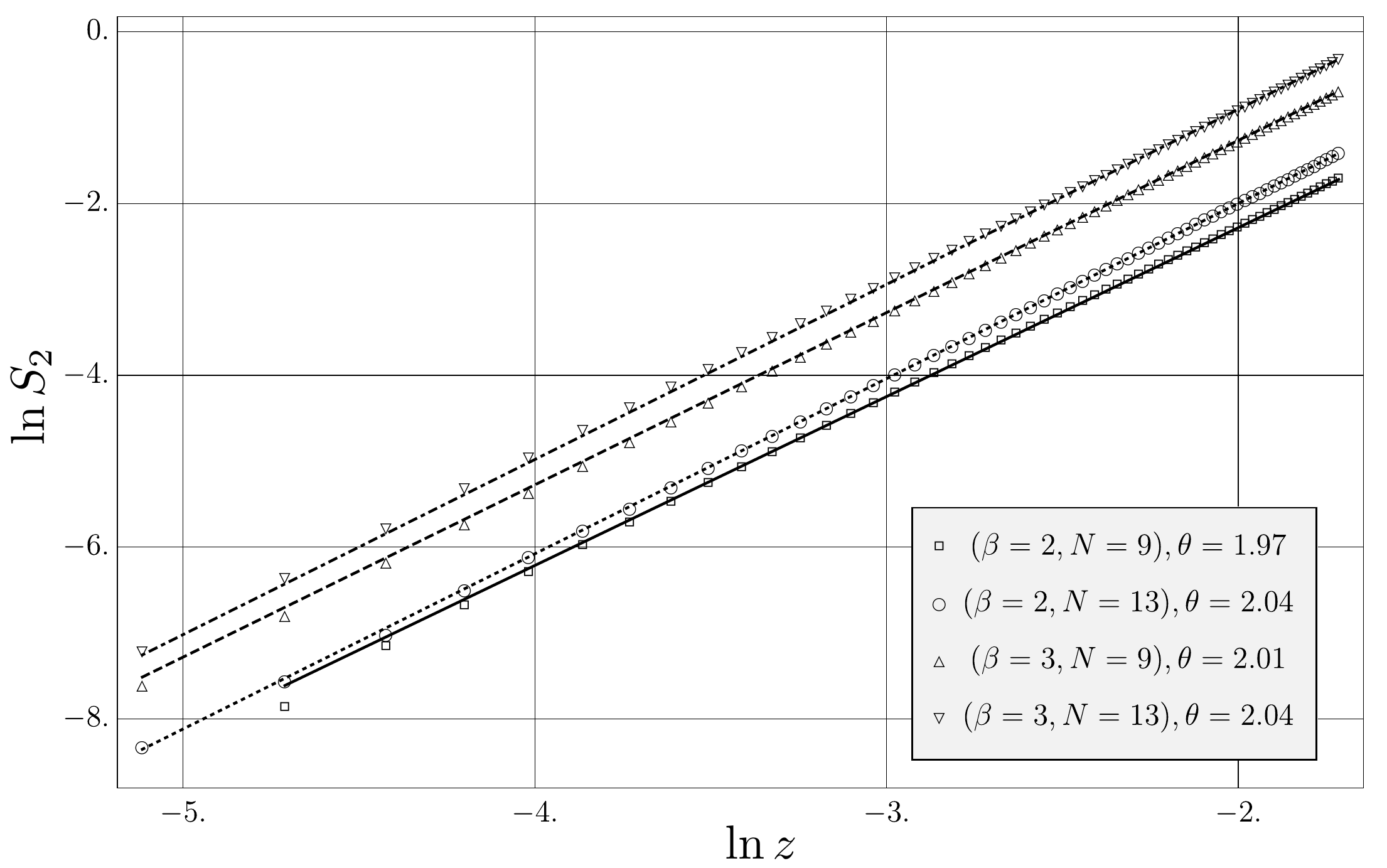}
\caption{Power-law behavior of the entanglement entropy, $S_2\sim \ell^\theta$,  for different values of  $\beta$ and $N$ and $a\to a_c$.}
\label{fig:fpowerlaw}
\end{figure}

\begin{figure}[t]
\includegraphics[width=0.9\linewidth]{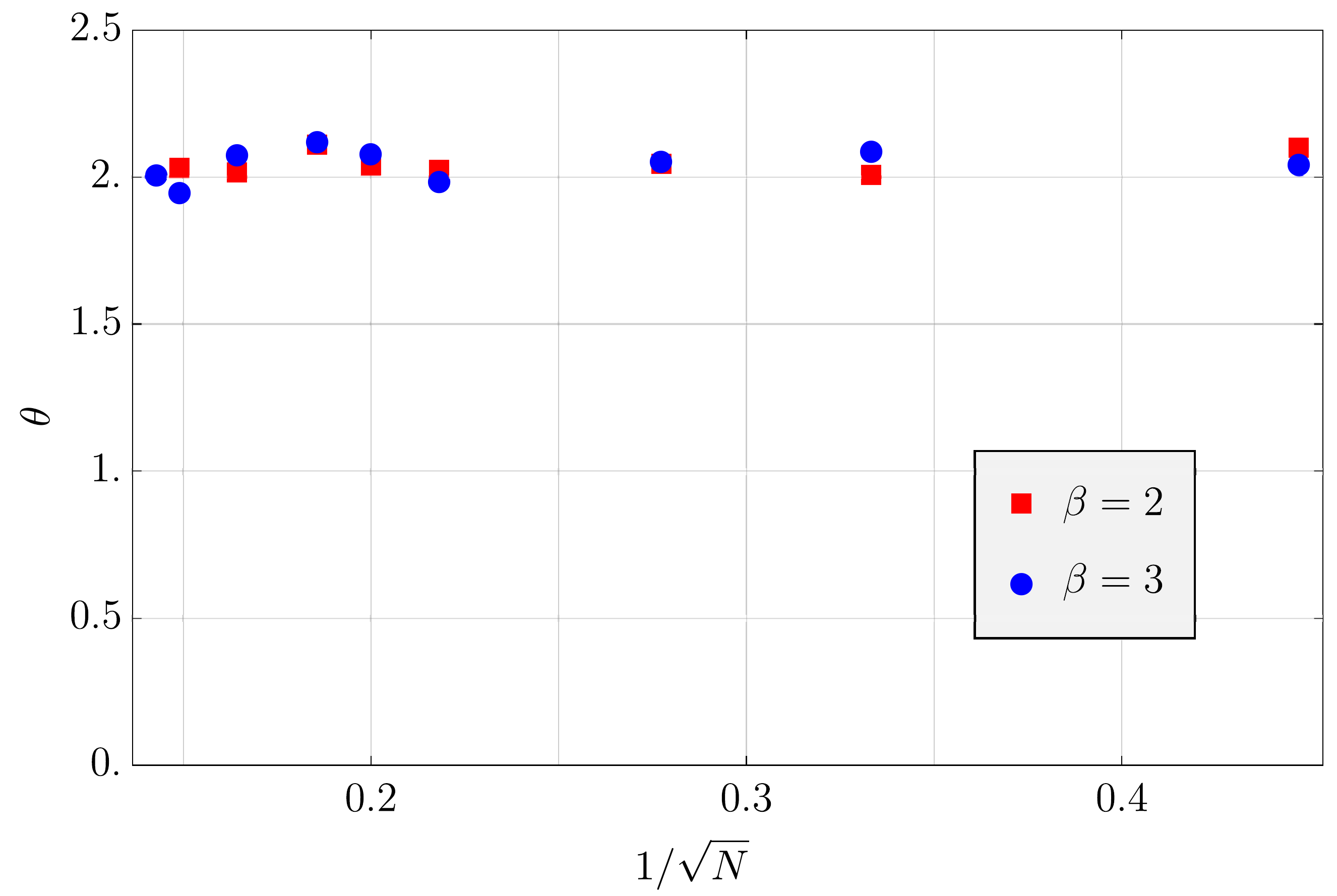}
\caption{The scaling exponent $\theta$ of the R\'enyi entropy as a function of the system size and for different values of the backflow exponent $\beta$.}
\label{fig:scalingExponent}
\end{figure}

In Fig.~\ref{fig:fpowerlaw}, we show the power-law behavior $S_2\sim\ell^\theta$ for different particle numbers ($N=9$ and $N=13$) and backflow exponents ($\beta=2$ and $\beta=3$). We find that the power laws 
are robust over a significant range with exponents that are very close to volume scaling, $\theta=2$. The deviations from a power law at very small scales are due to the small-distance cut-off $r_0$ of the backflow, Eq.~(\ref{eq:backflow}).
In Fig.~\ref{fig:scalingExponent} the extracted entanglement scaling exponents $\theta$ are shown for various particle numbers $N$. Our results show
that $\theta$ is \emph{independent} of $N$ and equal to $\theta=2$ within the error bars. Finally, we can change the backflow exponent from $\beta=3$ to $\beta=2$ which gives rise to a significant change of $d_H$ but does not affect the volume 
scaling of $S_2$.

\section{Discussion and conclusions}
\label{sec:disc}

Let us reiterate our main findings.  (i) The bipartite entanglement entropy is sensitive to the crossover between fractal and smooth behavior of the nodal structure. (ii) It is incapable to resolve the precise nature of the non-trivial fermion-sign structure in the fractal regime. We generically find here volume entanglement scaling, irrespective of the fractal dimension of the nodal structure. 

The findings we have presented in this work do fit into a broader development. Although the bipartite entanglement entropies have played a stimulating role with regard to the introduction of quantum information notions in quantum many-body- and field theory the realization is growing that it has severe shortcomings, to the degree that it may be plainly misleading. A first precedence has been the demonstration that it falls short of even detecting the quantum critical state of the transversal field Ising model in 2+1D \cite{Chandran:2013zqa}. The work by Grover and Fisher \cite{Grover:2014yea} that formed the initial motivation of the present study is also devastating: the gross scaling properties of the bipartite entropies are insensitive to the infinite party entanglement realized in configuration space, while the area law is just generic for a state characterized by a sign-free state where all wave function amplitudes are positive definite. Given this observation it appears to us that the widespread belief in the quantum information community that the ground states of systems described by any local Hamiltonian should show an area scaling of the bipartite entropy is actually based on folklore. Typically the focus has been on ground states of systems that can be enumerated, like the 1+1D systems and higher dimensional incompressible spin systems which are invariably sign-free. The (fermion) sign problem is just in the way of explicitly enumerating the ground states of sign-full systems and these have been just entirely ignored. These should exhibit a longer ranged entanglement entropy,  the simple case in point being the Fermi-gas with its area-log-area scaling.
Given the sensitivity of the bipartite entanglement entropies to the presence of signs, the next question to ask is whether the bipartite entropies are sensitive to any special features in the sign structure. This we set out to investigate in the this work. Given that basic scaling behavior is at stake, we perceive a sign structure that is organized as a fractal as an optimally beneficial circumstance for the R\'enyi entropy to reveal such specific sign information. The disappointing outcome is that the R\'enyi entropy appears to be only capable to discriminate between the Fermi-gas (with the area-log-area scaling) and the more dense sign structures characterized by fractal nodal surfaces. Although we did not check it explicitly, the odds are that for a nodal surface with {\em any} Hausdorff dimension $d_H$, $S_2$ will exhibit a volume scaling.

A clear indication is the extremely rapid convergence of $S_2$ as function of system size. As we emphasized in the previous section, the correct scaling behavior of the thermodynamic limit is already revealed departing from a system containing as few as five fermions. This is in stark contrast with the geometry of the nodal surface itself where one needs at least 10 times as many particles to discern the fractal dimension, overcoming the finite size artifacts. Obviously, this signals that the bipartite entanglement entropy, even in the presence of signs, is just revealing short distance information. The precise mechanism is unclear to us. 

Arguably, the only genuine non-Fermi liquids that have been identified based on controlled mathematics are the holographic strange metals. As we explained in the introduction their vacuum state  may be characterized by bipartite von Neumann entropies with anomalous dimensions intermediate between area and volume scaling. However, this scaling dimension coincides with the  hyperscaling violation exponent $\theta$ that is well established to characterize the deep infrared \cite{zaanen2015holographic}. Given the message of the previous paragraph this appears as an apparent paradox: how can it be that a quantity (the entanglement entropy) that appears to only pick up UV information in an explicit field theoretical setting (the back-flow fermions) manages to measure a deep IR scaling dimension (the $\theta$ of the holographic strange metals)? The caveat is of course that the first quantized back-flow system which is in fact based on a hidden free fermion ansatz cannot be assigned a universal status being representative for the sign structure of all quantum critical fermion systems. We leave the origin of the anomalous scaling of the entanglement entropy of the holographic strange metals as the great challenge for future research.

Given that the bipartite entanglement entropies are such blunt tools dealing with the complexities of sign-full quantum states of matter, are there alternatives? It seems obvious that sign structure and infinite party, field theoretical entanglement go hand in hand. This is generally recognized dealing with highly excited states. This raises then the question whether perhaps the nodal surface might serve the purpose of further characterizing the nature of this entanglement. 

It has the obvious disadvantage that it requires too much information. In principle the full density matrix contains all quantum information of the state under consideration but it is just too complex. The nodal surface is defined as the hyper-surface of zero's of this density matrix, amounting to only a rather marginal improvement with regard to the information overload. However, the advantage of the nodal surface is that the sign structure is {\em geometrized}.  A case in point is the very notion that the nodal surface can be characterized by either a smooth or a fractal geometry. In fact, because of the difficulty that one needs the state explicitly in order to enumerate the nodal surface, very little is known regarding this nodal surface geometry. A next difficulty is that the relationship between the nodal surface and the infinite party entanglement is far from straightforward. Although not quite enumerated, the quantum non-locality associated with the antisymmetrized states in the Fock space of the free fermion gas has a mirror image in the non-locality of the Fermi gas nodal surface. Changing the position of one particle will change the precise locus of the nodal surface everywhere else. Obviously, turning the smooth nodal surface geometry into the fractal one of the back-flow system this degree of non-locality is further enhanced and one could then argue that the
quantum critical fermion system is more densely entangled.

However, the lesson of the Bell pairs is that entanglement should be representation independent. Although it involves a non-local and highly singular transformation, the back-flow system can be represented as a free fermion gas of ``back-flow particles'' and the denser entanglement associated with the bare particle coordinates can therefore be viewed as a peculiarity of an inconvenient representation. Even for the Fermi gas itself there is a confusing issue with the precise status of the ``antisymmetrization entanglement''. The permutation signs surely block the way to a short range entangled product state in real space. However, real space is a choice of representation. How does this work in single particle momentum space? In fact, using the constrained path integral one can prove easily that the Fermi-gas precisely maps on the problem of a classical Mott insulator living in a harmonic well in momentum space \cite{2008arXiv0802.2455Z}. Although this ``Mottness'' involves non-local information this is the same kind of non-locality perceived by a car stuck in a traffic jam: this has no relation whatever with the quantum information that may be used to factorize primes in polynomial time.

So much is clear that the fermion-sign problem is coincident with the present incapacity of the available computational tools to deal with infinite party long range entangled states of thermodynamically large systems. But this incapacity does not necessarily imply the end of physics: especially dealing with the incomputable sign-full vacuum states there should be physics which may quite well be beautiful physics as suggested by the holographic strange metals.  An uncharted territory of physics is waiting behind the fermion-sign brick wall and we hope that our investigations may stimulate others to have a closer look.

\acknowledgments{
This work was carried out on the Dutch national e-infrastructure with the support of SURF Foundation. We are thankful to Tarun Grover who suggested to us having a look at the R\'enyi entropy of the 
backflow system and pointing out the algorithm explained in Sec.~\ref{sec:Renyi}.  We acknowledge support of a grant from the John Templeton Foundation. The opinions expressed in this publication are those of the authors and do not necessarily reflect the views of the John Templeton Foundation.}

\appendix

\section{Computing the R\'enyi Entropy: the algorithm}
\label{sec:Renyi}

In this section we will review  the Monte-Carlo algorithm \cite{PhysRevLett.107.067202,Hastings+10} to calculate the second R\'enyi entropy $S_2$ for any given normalized $N$-particle wavefunction $\Psi(\bs{r}_1,\ldots,\bs{r}_N)$. This algorithm will then be used in Sec.~\ref{sec:results} to calculate $S_2$ for the fermionic backflow wavefunctions (\ref{eq:waveFunction}). Notice that we expect $S_{vN}$ and $S_n$ to have the same scaling properties. $S_2$ is chosen because it is relatively easy to calculate. 

In order to compute the bipartite entanglement entropy one needs to split the system into two sub-systems $A$ and $B$, as illustrated in Fig.~\ref{fig:entanglementArea}. For brevity, we write the wavefunction as $\Psi(\bs{r}_1,\ldots,\bs{r}_N) = \Psi (\alpha,\beta)$ where $\alpha$ and $\beta$ are the configurations of the sub-system $A$ and $B$, respectively (e.g. $\alpha$ is determined by the number $N_A$ of particles in sub-system $A$ and by the positions $\bs{r}_1,\ldots, \bs{r}_{N_A}$ of these particles and $N=N_A+N_B$). The reduced density matrix can then be written as 
\begin{equation}
\rho_A (\alpha, \alpha') = \sum_{\beta}\Psi^*(\alpha,\beta) \Psi(\alpha',\beta),
\end{equation}
where the partial trace over sub-system $B$ implies that $N_A=N_A'$. This however does not fix $N_A$ to be a particular number. One should view the above definition as a short-hand notation for
\begin{eqnarray}
\rho_A (\alpha, \alpha') & = &  \delta_{N_A,0}+\delta_{N_A,1}\rho_A^{(1)}(\bs{r}_1;\bs{r}_1')+\ldots\nonumber\\
& & +\delta_{N_A,N} \rho_A^{(N)} (\bs{r}_1,\ldots\bs{r}_N;\bs{r}_1',\ldots,\bs{r}_N').
\end{eqnarray}
The exponentiated 2nd R\'enyi entanglement entropy $e^{-S_2} = \textrm{Tr}\hat{\rho}_A^2$ can then be expressed as
\begin{equation}
e^{-S_2} = \sum_{\alpha\alpha' \beta\beta'}\Psi^*(\alpha,\beta)\Psi(\alpha',\beta)\Psi^*(\alpha',\beta')\Psi(\alpha,\beta').
\label{eq:renyiEntropy}
\end{equation}
This expression has a simple physical interpretation. It is equal to the expectation value $e^{-S_2} = \bra \hat{\textrm{Swap}}_B \ket $ of a swap operator that exchanges the configurations of sub-system $B$ between two replicas $\mathcal{S}$, $\mathcal{S}'$ of the system, $\hat{\textrm{Swap}}_B |\alpha,\beta\ket |\alpha',\beta'\ket = |\alpha,\beta'\ket |\alpha',\beta\ket$.

Note that the sum in Eq.~(\ref{eq:renyiEntropy}) is equivalent to an integral over the positions $\bs{r}_1,\ldots,\bs{r}_N$ in $\mathcal{S}$ and $\bs{r}_1',\ldots,\bs{r}_N'$ in $\mathcal{S}'$. This high-dimensional integral we will calculate numerically, using Metropolis Monte-Carlo integration. We will sample configurations using the probability distribution 
\begin{eqnarray}
P(\alpha,\beta;\alpha',\beta') & = &  |\Psi(\alpha,\beta)|^2\cdot |\Psi(\alpha',\beta')|^2\nonumber \\
& = &  |\Psi(\bs{r}_1,\ldots\bs{r}_N)|^2\cdot  |\Psi(\bs{r}_1',\ldots,\bs{r}_N')|^2.
\end{eqnarray}
After a simple re-writing of the integrand in Eq.~(\ref{eq:renyiEntropy}), we find that we need to average
\begin{equation}
F(\alpha,\beta;\alpha',\beta') = \frac{\Psi(\alpha',\beta) \Psi(\alpha,\beta')}{\Psi(\alpha,\beta) \Psi(\alpha',\beta')}
\end{equation}
over Markov chains generated with the probability distribution $P$.


\begin{thebibliography}{51}
\expandafter\ifx\csname natexlab\endcsname\relax\def\natexlab#1{#1}\fi
\expandafter\ifx\csname bibnamefont\endcsname\relax
  \def\bibnamefont#1{#1}\fi
\expandafter\ifx\csname bibfnamefont\endcsname\relax
  \def\bibfnamefont#1{#1}\fi
\expandafter\ifx\csname citenamefont\endcsname\relax
  \def\citenamefont#1{#1}\fi
\expandafter\ifx\csname url\endcsname\relax
  \def\url#1{\texttt{#1}}\fi
\expandafter\ifx\csname urlprefix\endcsname\relax\def\urlprefix{URL }\fi
\providecommand{\bibinfo}[2]{#2}
\providecommand{\eprint}[2][]{\url{#2}}

\bibitem[{\citenamefont{Laflorencie}(2016)}]{Laflorencie:2015eck}
\bibinfo{author}{\bibfnamefont{N.}~\bibnamefont{Laflorencie}},
  \bibinfo{journal}{Physics Reports} \textbf{\bibinfo{volume}{646}},
  \bibinfo{pages}{1} (\bibinfo{year}{2016}).

\bibitem[{\citenamefont{Bombelli et~al.}(1986)\citenamefont{Bombelli, Koul,
  Lee, and Sokin}}]{Bombelli+86}
\bibinfo{author}{\bibfnamefont{L.}~\bibnamefont{Bombelli}},
  \bibinfo{author}{\bibfnamefont{R.~K.} \bibnamefont{Koul}},
  \bibinfo{author}{\bibfnamefont{J.}~\bibnamefont{Lee}}, \bibnamefont{and}
  \bibinfo{author}{\bibfnamefont{R.~D.} \bibnamefont{Sokin}},
  \bibinfo{journal}{Phys. Rev. D} \textbf{\bibinfo{volume}{34}},
  \bibinfo{pages}{373} (\bibinfo{year}{1986}).

\bibitem[{\citenamefont{Srednicki}(1993)}]{Srednicki93}
\bibinfo{author}{\bibfnamefont{M.}~\bibnamefont{Srednicki}},
  \bibinfo{journal}{Phys. Rev. Lett.} \textbf{\bibinfo{volume}{71}},
  \bibinfo{pages}{666} (\bibinfo{year}{1993}).

\bibitem[{\citenamefont{Callan and Wilczek}(1994)}]{Callan+94}
\bibinfo{author}{\bibfnamefont{C.}~\bibnamefont{Callan}} \bibnamefont{and}
  \bibinfo{author}{\bibfnamefont{F.}~\bibnamefont{Wilczek}},
  \bibinfo{journal}{Physic Letters B} \textbf{\bibinfo{volume}{55}},
  \bibinfo{pages}{333} (\bibinfo{year}{1994}).

\bibitem[{\citenamefont{Kallin et~al.}(2011)\citenamefont{Kallin, Hastings,
  Melko, and Singh}}]{Kallin+11}
\bibinfo{author}{\bibfnamefont{A.~B.} \bibnamefont{Kallin}},
  \bibinfo{author}{\bibfnamefont{M.~B.} \bibnamefont{Hastings}},
  \bibinfo{author}{\bibfnamefont{R.~G.} \bibnamefont{Melko}}, \bibnamefont{and}
  \bibinfo{author}{\bibfnamefont{R.~R.~P.} \bibnamefont{Singh}},
  \bibinfo{journal}{Phys. Rev. B} \textbf{\bibinfo{volume}{84}},
  \bibinfo{pages}{165134} (\bibinfo{year}{2011}).

\bibitem[{\citenamefont{Furukawa and Misguich}(2007)}]{Furukawa+07}
\bibinfo{author}{\bibfnamefont{S.}~\bibnamefont{Furukawa}} \bibnamefont{and}
  \bibinfo{author}{\bibfnamefont{G.}~\bibnamefont{Misguich}},
  \bibinfo{journal}{Phys. Rev. B} \textbf{\bibinfo{volume}{75}},
  \bibinfo{pages}{214407} (\bibinfo{year}{2007}).

\bibitem[{\citenamefont{Metlitski et~al.}(2009)\citenamefont{Metlitski,
  Fuertes, and Sachdev}}]{Metlitski+09}
\bibinfo{author}{\bibfnamefont{M.~A.} \bibnamefont{Metlitski}},
  \bibinfo{author}{\bibfnamefont{C.~A.} \bibnamefont{Fuertes}},
  \bibnamefont{and} \bibinfo{author}{\bibfnamefont{S.}~\bibnamefont{Sachdev}},
  \bibinfo{journal}{Phys. Rev. B} \textbf{\bibinfo{volume}{80}},
  \bibinfo{pages}{115122} (\bibinfo{year}{2009}).

\bibitem[{\citenamefont{Helmes and Wessel}(2014)}]{Helmes+14}
\bibinfo{author}{\bibfnamefont{J.}~\bibnamefont{Helmes}} \bibnamefont{and}
  \bibinfo{author}{\bibfnamefont{S.}~\bibnamefont{Wessel}},
  \bibinfo{journal}{Phys. Rev. B} \textbf{\bibinfo{volume}{89}},
  \bibinfo{pages}{245120} (\bibinfo{year}{2014}).

\bibitem[{\citenamefont{Fr\'erot and Roscilde}(2016)}]{Frerot+16}
\bibinfo{author}{\bibfnamefont{I.}~\bibnamefont{Fr\'erot}} \bibnamefont{and}
  \bibinfo{author}{\bibfnamefont{T.}~\bibnamefont{Roscilde}},
  \bibinfo{journal}{Phys. Rev. Lett.} \textbf{\bibinfo{volume}{116}},
  \bibinfo{pages}{190401} (\bibinfo{year}{2016}).

\bibitem[{\citenamefont{Chandran et~al.}(2014)\citenamefont{Chandran, Khemani,
  and Sondhi}}]{Chandran:2013zqa}
\bibinfo{author}{\bibfnamefont{A.}~\bibnamefont{Chandran}},
  \bibinfo{author}{\bibfnamefont{V.}~\bibnamefont{Khemani}}, \bibnamefont{and}
  \bibinfo{author}{\bibfnamefont{S.~L.} \bibnamefont{Sondhi}},
  \bibinfo{journal}{Phys. Rev. Lett.} \textbf{\bibinfo{volume}{113}},
  \bibinfo{pages}{060501} (\bibinfo{year}{2014}).

\bibitem[{\citenamefont{Page}(1993)}]{Page93}
\bibinfo{author}{\bibfnamefont{D.~N.} \bibnamefont{Page}},
  \bibinfo{journal}{Phys. Rev. Lett.} \textbf{\bibinfo{volume}{71}},
  \bibinfo{pages}{1291} (\bibinfo{year}{1993}).

\bibitem[{\citenamefont{Foong and Kanno}(1994)}]{Foong+94}
\bibinfo{author}{\bibfnamefont{S.~K.} \bibnamefont{Foong}} \bibnamefont{and}
  \bibinfo{author}{\bibfnamefont{S.}~\bibnamefont{Kanno}},
  \bibinfo{journal}{Phys. Rev. Lett.} \textbf{\bibinfo{volume}{72}},
  \bibinfo{pages}{1148} (\bibinfo{year}{1994}).

\bibitem[{\citenamefont{Sanchez-Ruiz}(1995)}]{Sanchez95}
\bibinfo{author}{\bibfnamefont{J.}~\bibnamefont{Sanchez-Ruiz}},
  \bibinfo{journal}{Phys. Rev. E} \textbf{\bibinfo{volume}{52}},
  \bibinfo{pages}{5653} (\bibinfo{year}{1995}).

\bibitem[{\citenamefont{Sen}(1996)}]{Sen96}
\bibinfo{author}{\bibfnamefont{S.}~\bibnamefont{Sen}}, \bibinfo{journal}{Phys.
  Rev. Lett.} \textbf{\bibinfo{volume}{77}}, \bibinfo{pages}{1}
  (\bibinfo{year}{1996}).

\bibitem[{\citenamefont{Grover and Fisher}(2014)}]{Grover:2013tia}
\bibinfo{author}{\bibfnamefont{T.}~\bibnamefont{Grover}} \bibnamefont{and}
  \bibinfo{author}{\bibfnamefont{M.~P.~A.} \bibnamefont{Fisher}},
  \bibinfo{journal}{J. Stat. Mech.} \textbf{\bibinfo{volume}{1410}},
  \bibinfo{pages}{P10010} (\bibinfo{year}{2014}).

\bibitem[{\citenamefont{Grover and Fisher}(2015)}]{Grover:2014yea}
\bibinfo{author}{\bibfnamefont{T.}~\bibnamefont{Grover}} \bibnamefont{and}
  \bibinfo{author}{\bibfnamefont{M.~P.~A.} \bibnamefont{Fisher}},
  \bibinfo{journal}{Phys. Rev. A} \textbf{\bibinfo{volume}{92}},
  \bibinfo{pages}{042308} (\bibinfo{year}{2015}).

\bibitem[{\citenamefont{Feynman}(1972)}]{Feynman72}
\bibinfo{author}{\bibfnamefont{R.~P.} \bibnamefont{Feynman}},
  \emph{\bibinfo{title}{Statistical Mechanics, A Set of Lectures}}
  (\bibinfo{publisher}{Addison-Wesley Publishing Company},
  \bibinfo{year}{1972}).

\bibitem[{\citenamefont{Wu}(2009)}]{Wu09}
\bibinfo{author}{\bibfnamefont{C.}~\bibnamefont{Wu}}, \bibinfo{journal}{Modern
  Physics Letters B} \textbf{\bibinfo{volume}{23}}, \bibinfo{pages}{1}
  (\bibinfo{year}{2009}).

\bibitem[{\citenamefont{Troyer and Wiese}(2005)}]{Troyer:2004ge}
\bibinfo{author}{\bibfnamefont{M.}~\bibnamefont{Troyer}} \bibnamefont{and}
  \bibinfo{author}{\bibfnamefont{U.-J.} \bibnamefont{Wiese}},
  \bibinfo{journal}{Phys. Rev. Lett.} \textbf{\bibinfo{volume}{94}},
  \bibinfo{pages}{170201} (\bibinfo{year}{2005}).

\bibitem[{\citenamefont{Wolf}(2006)}]{PhysRevLett.96.010404}
\bibinfo{author}{\bibfnamefont{M.~M.} \bibnamefont{Wolf}},
  \bibinfo{journal}{Phys. Rev. Lett.} \textbf{\bibinfo{volume}{96}},
  \bibinfo{pages}{010404} (\bibinfo{year}{2006}).

\bibitem[{\citenamefont{Gioev and Klich}(2006)}]{PhysRevLett.96.100503}
\bibinfo{author}{\bibfnamefont{D.}~\bibnamefont{Gioev}} \bibnamefont{and}
  \bibinfo{author}{\bibfnamefont{I.}~\bibnamefont{Klich}},
  \bibinfo{journal}{Phys. Rev. Lett.} \textbf{\bibinfo{volume}{96}},
  \bibinfo{pages}{100503} (\bibinfo{year}{2006}).

\bibitem[{\citenamefont{Li et~al.}(2006)\citenamefont{Li, Ding, Yu, Roscilde,
  and Haas}}]{PhysRevB.74.073103}
\bibinfo{author}{\bibfnamefont{W.}~\bibnamefont{Li}},
  \bibinfo{author}{\bibfnamefont{L.}~\bibnamefont{Ding}},
  \bibinfo{author}{\bibfnamefont{R.}~\bibnamefont{Yu}},
  \bibinfo{author}{\bibfnamefont{T.}~\bibnamefont{Roscilde}}, \bibnamefont{and}
  \bibinfo{author}{\bibfnamefont{S.}~\bibnamefont{Haas}},
  \bibinfo{journal}{Phys. Rev. B} \textbf{\bibinfo{volume}{74}},
  \bibinfo{pages}{073103} (\bibinfo{year}{2006}).

\bibitem[{\citenamefont{Barthel et~al.}(2006)\citenamefont{Barthel, Chung, and
  Schollw\"ock}}]{PhysRevA.74.022329}
\bibinfo{author}{\bibfnamefont{T.}~\bibnamefont{Barthel}},
  \bibinfo{author}{\bibfnamefont{M.-C.} \bibnamefont{Chung}}, \bibnamefont{and}
  \bibinfo{author}{\bibfnamefont{U.}~\bibnamefont{Schollw\"ock}},
  \bibinfo{journal}{Phys. Rev. A} \textbf{\bibinfo{volume}{74}},
  \bibinfo{pages}{022329} (\bibinfo{year}{2006}).

\bibitem[{\citenamefont{McMinis and Tubman}(2013)}]{PhysRevB.87.081108}
\bibinfo{author}{\bibfnamefont{J.}~\bibnamefont{McMinis}} \bibnamefont{and}
  \bibinfo{author}{\bibfnamefont{N.~M.} \bibnamefont{Tubman}},
  \bibinfo{journal}{Phys. Rev. B} \textbf{\bibinfo{volume}{87}},
  \bibinfo{pages}{081108} (\bibinfo{year}{2013}).

\bibitem[{bos()}]{bose_metal}
\bibinfo{note}{This behavior can yet be understood in a bosonic setting, be it
  that the corresponding boson problem is pathological. The area-log-area
  behavior is exhibited by any system carrying a spectrum of excitations
  characterized by a surface of dimension $d-1$ in single particle momentum
  space of massless excitations which are governed at any point of this surface
  by a CFT$^2$ (scale invariant system in 1+1D). This is just an alternative
  way of naming a Fermi surface
  \cite{PhysRevLett.105.050502,PhysRevB.86.035116} while also a highly fine
  tuned frustrated Bose system can be tailored that it is characterized by a
  ``'Bose surface'' of gapless excitations at its quantum critical point
  \cite{PhysRevB.66.054526}. For such ``'Bose metals'' a logarithmic
  enhancement of the area law was found
  \cite{PhysRevLett.111.210402,PhysRevB.93.121109}. Self-evidently, gapless
  quantum spin liquids with a spinor Fermi surface also show a logarithmic
  correction to the area law \cite{PhysRevLett.107.067202}.}

\bibitem[{\citenamefont{Zaanen et~al.}(2015)\citenamefont{Zaanen, Liu, Sun, and
  Schalm}}]{zaanen2015holographic}
\bibinfo{author}{\bibfnamefont{J.}~\bibnamefont{Zaanen}},
  \bibinfo{author}{\bibfnamefont{Y.}~\bibnamefont{Liu}},
  \bibinfo{author}{\bibfnamefont{Y.}~\bibnamefont{Sun}}, \bibnamefont{and}
  \bibinfo{author}{\bibfnamefont{K.}~\bibnamefont{Schalm}},
  \emph{\bibinfo{title}{Holographic Duality in Condensed Matter Physics}}
  (\bibinfo{publisher}{Cambridge University Press}, \bibinfo{year}{2015}), ISBN
  \bibinfo{isbn}{9781107080089}.

\bibitem[{\citenamefont{Huijse et~al.}(2012)\citenamefont{Huijse, Sachdev, and
  Swingle}}]{PhysRevB.85.035121}
\bibinfo{author}{\bibfnamefont{L.}~\bibnamefont{Huijse}},
  \bibinfo{author}{\bibfnamefont{S.}~\bibnamefont{Sachdev}}, \bibnamefont{and}
  \bibinfo{author}{\bibfnamefont{B.}~\bibnamefont{Swingle}},
  \bibinfo{journal}{Phys. Rev. B} \textbf{\bibinfo{volume}{85}},
  \bibinfo{pages}{035121} (\bibinfo{year}{2012}).

\bibitem[{\citenamefont{Fisher}(1986)}]{PhysRevLett.56.416}
\bibinfo{author}{\bibfnamefont{D.~S.} \bibnamefont{Fisher}},
  \bibinfo{journal}{Phys. Rev. Lett.} \textbf{\bibinfo{volume}{56}},
  \bibinfo{pages}{416} (\bibinfo{year}{1986}).

\bibitem[{\citenamefont{Schmidt et~al.}(1981)\citenamefont{Schmidt, Lee, Kalos,
  and Chester}}]{PhysRevLett.47.807}
\bibinfo{author}{\bibfnamefont{K.~E.} \bibnamefont{Schmidt}},
  \bibinfo{author}{\bibfnamefont{M.~A.} \bibnamefont{Lee}},
  \bibinfo{author}{\bibfnamefont{M.~H.} \bibnamefont{Kalos}}, \bibnamefont{and}
  \bibinfo{author}{\bibfnamefont{G.~V.} \bibnamefont{Chester}},
  \bibinfo{journal}{Phys. Rev. Lett.} \textbf{\bibinfo{volume}{47}},
  \bibinfo{pages}{807} (\bibinfo{year}{1981}).

\bibitem[{\citenamefont{Kwon et~al.}(1993)\citenamefont{Kwon, Ceperley, and
  Martin}}]{PhysRevB.48.12037}
\bibinfo{author}{\bibfnamefont{Y.}~\bibnamefont{Kwon}},
  \bibinfo{author}{\bibfnamefont{D.~M.} \bibnamefont{Ceperley}},
  \bibnamefont{and} \bibinfo{author}{\bibfnamefont{R.~M.}
  \bibnamefont{Martin}}, \bibinfo{journal}{Phys. Rev. B}
  \textbf{\bibinfo{volume}{48}}, \bibinfo{pages}{12037} (\bibinfo{year}{1993}).

\bibitem[{\citenamefont{Kwon et~al.}(1998)\citenamefont{Kwon, Ceperley, and
  Martin}}]{PhysRevB.58.6800}
\bibinfo{author}{\bibfnamefont{Y.}~\bibnamefont{Kwon}},
  \bibinfo{author}{\bibfnamefont{D.~M.} \bibnamefont{Ceperley}},
  \bibnamefont{and} \bibinfo{author}{\bibfnamefont{R.~M.}
  \bibnamefont{Martin}}, \bibinfo{journal}{Phys. Rev. B}
  \textbf{\bibinfo{volume}{58}}, \bibinfo{pages}{6800} (\bibinfo{year}{1998}).

\bibitem[{\citenamefont{Holzmann et~al.}(2003)\citenamefont{Holzmann, Ceperley,
  Pierleoni, and Esler}}]{PhysRevE.68.046707}
\bibinfo{author}{\bibfnamefont{M.}~\bibnamefont{Holzmann}},
  \bibinfo{author}{\bibfnamefont{D.~M.} \bibnamefont{Ceperley}},
  \bibinfo{author}{\bibfnamefont{C.}~\bibnamefont{Pierleoni}},
  \bibnamefont{and} \bibinfo{author}{\bibfnamefont{K.}~\bibnamefont{Esler}},
  \bibinfo{journal}{Phys. Rev. E} \textbf{\bibinfo{volume}{68}},
  \bibinfo{pages}{046707} (\bibinfo{year}{2003}).

\bibitem[{\citenamefont{Tocchio et~al.}(2008)\citenamefont{Tocchio, Becca,
  Parola, and Sorella}}]{PhysRevB.78.041101}
\bibinfo{author}{\bibfnamefont{L.~F.} \bibnamefont{Tocchio}},
  \bibinfo{author}{\bibfnamefont{F.}~\bibnamefont{Becca}},
  \bibinfo{author}{\bibfnamefont{A.}~\bibnamefont{Parola}}, \bibnamefont{and}
  \bibinfo{author}{\bibfnamefont{S.}~\bibnamefont{Sorella}},
  \bibinfo{journal}{Phys. Rev. B} \textbf{\bibinfo{volume}{78}},
  \bibinfo{pages}{041101} (\bibinfo{year}{2008}).

\bibitem[{\citenamefont{{Kr{\"u}ger} and {Zaanen}}(2008)}]{2008PhRvB..78c5104K}
\bibinfo{author}{\bibfnamefont{F.}~\bibnamefont{{Kr{\"u}ger}}}
  \bibnamefont{and} \bibinfo{author}{\bibfnamefont{J.}~\bibnamefont{{Zaanen}}},
  \bibinfo{journal}{\prb} \textbf{\bibinfo{volume}{78}}, \bibinfo{eid}{035104}
  (\bibinfo{year}{2008}).

\bibitem[{\citenamefont{Ceperley}(1991)}]{Ceperley91}
\bibinfo{author}{\bibfnamefont{D.~M.} \bibnamefont{Ceperley}},
  \bibinfo{journal}{Journal of Statistical Physics}
  \textbf{\bibinfo{volume}{63}}, \bibinfo{pages}{1237} (\bibinfo{year}{1991}),
  ISSN \bibinfo{issn}{1572-9613}.

\bibitem[{\citenamefont{Ceperley}(1992)}]{PhysRevLett.69.331}
\bibinfo{author}{\bibfnamefont{D.~M.} \bibnamefont{Ceperley}},
  \bibinfo{journal}{Phys. Rev. Lett.} \textbf{\bibinfo{volume}{69}},
  \bibinfo{pages}{331} (\bibinfo{year}{1992}).

\bibitem[{\citenamefont{Pierleoni et~al.}(1994)\citenamefont{Pierleoni,
  Ceperley, Bernu, and Magro}}]{PhysRevLett.73.2145}
\bibinfo{author}{\bibfnamefont{C.}~\bibnamefont{Pierleoni}},
  \bibinfo{author}{\bibfnamefont{D.~M.} \bibnamefont{Ceperley}},
  \bibinfo{author}{\bibfnamefont{B.}~\bibnamefont{Bernu}}, \bibnamefont{and}
  \bibinfo{author}{\bibfnamefont{W.~R.} \bibnamefont{Magro}},
  \bibinfo{journal}{Phys. Rev. Lett.} \textbf{\bibinfo{volume}{73}},
  \bibinfo{pages}{2145} (\bibinfo{year}{1994}).

\bibitem[{\citenamefont{Magro et~al.}(1996)\citenamefont{Magro, Ceperley,
  Pierleoni, and Bernu}}]{PhysRevLett.76.1240}
\bibinfo{author}{\bibfnamefont{W.~R.} \bibnamefont{Magro}},
  \bibinfo{author}{\bibfnamefont{D.~M.} \bibnamefont{Ceperley}},
  \bibinfo{author}{\bibfnamefont{C.}~\bibnamefont{Pierleoni}},
  \bibnamefont{and} \bibinfo{author}{\bibfnamefont{B.}~\bibnamefont{Bernu}},
  \bibinfo{journal}{Phys. Rev. Lett.} \textbf{\bibinfo{volume}{76}},
  \bibinfo{pages}{1240} (\bibinfo{year}{1996}).

\bibitem[{\citenamefont{Marel et~al.}(2003)\citenamefont{Marel, Molegraaf,
  Zaanen, Nussinov, Carbone, Damascelli, Eisaki, Greven, Kes, and
  Li}}]{Marel+03}
\bibinfo{author}{\bibfnamefont{D.~v.~d.} \bibnamefont{Marel}},
  \bibinfo{author}{\bibfnamefont{H.~J.~A.} \bibnamefont{Molegraaf}},
  \bibinfo{author}{\bibfnamefont{J.}~\bibnamefont{Zaanen}},
  \bibinfo{author}{\bibfnamefont{Z.}~\bibnamefont{Nussinov}},
  \bibinfo{author}{\bibfnamefont{F.}~\bibnamefont{Carbone}},
  \bibinfo{author}{\bibfnamefont{A.}~\bibnamefont{Damascelli}},
  \bibinfo{author}{\bibfnamefont{H.}~\bibnamefont{Eisaki}},
  \bibinfo{author}{\bibfnamefont{M.}~\bibnamefont{Greven}},
  \bibinfo{author}{\bibfnamefont{P.~H.} \bibnamefont{Kes}}, \bibnamefont{and}
  \bibinfo{author}{\bibfnamefont{M.}~\bibnamefont{Li}},
  \bibinfo{journal}{Nature} \textbf{\bibinfo{volume}{425}},
  \bibinfo{pages}{271} (\bibinfo{year}{2003}).

\bibitem[{\citenamefont{{Paschen} et~al.}(2004)\citenamefont{{Paschen},
  {L{\"u}hmann}, {Wirth}, {Gegenwart}, {Trovarelli}, {Geibel}, {Steglich},
  {Coleman}, and {Si}}}]{2004Natur.432..881P}
\bibinfo{author}{\bibfnamefont{S.}~\bibnamefont{{Paschen}}},
  \bibinfo{author}{\bibfnamefont{T.}~\bibnamefont{{L{\"u}hmann}}},
  \bibinfo{author}{\bibfnamefont{S.}~\bibnamefont{{Wirth}}},
  \bibinfo{author}{\bibfnamefont{P.}~\bibnamefont{{Gegenwart}}},
  \bibinfo{author}{\bibfnamefont{O.}~\bibnamefont{{Trovarelli}}},
  \bibinfo{author}{\bibfnamefont{C.}~\bibnamefont{{Geibel}}},
  \bibinfo{author}{\bibfnamefont{F.}~\bibnamefont{{Steglich}}},
  \bibinfo{author}{\bibfnamefont{P.}~\bibnamefont{{Coleman}}},
  \bibnamefont{and} \bibinfo{author}{\bibfnamefont{Q.}~\bibnamefont{{Si}}},
  \bibinfo{journal}{\nat} \textbf{\bibinfo{volume}{432}}, \bibinfo{pages}{881}
  (\bibinfo{year}{2004}).

\bibitem[{\citenamefont{Custers et~al.}(2003)\citenamefont{Custers, Gegenwart,
  Wilhelm, Neumaier, Tokiwa, Trovarelli, Geibel, Steglich, Pepin, and
  Coleman}}]{Custers+03}
\bibinfo{author}{\bibfnamefont{J.}~\bibnamefont{Custers}},
  \bibinfo{author}{\bibfnamefont{P.}~\bibnamefont{Gegenwart}},
  \bibinfo{author}{\bibfnamefont{H.}~\bibnamefont{Wilhelm}},
  \bibinfo{author}{\bibfnamefont{K.}~\bibnamefont{Neumaier}},
  \bibinfo{author}{\bibfnamefont{Y.}~\bibnamefont{Tokiwa}},
  \bibinfo{author}{\bibfnamefont{O.}~\bibnamefont{Trovarelli}},
  \bibinfo{author}{\bibfnamefont{C.}~\bibnamefont{Geibel}},
  \bibinfo{author}{\bibfnamefont{F.}~\bibnamefont{Steglich}},
  \bibinfo{author}{\bibfnamefont{C.}~\bibnamefont{Pepin}}, \bibnamefont{and}
  \bibinfo{author}{\bibfnamefont{P.}~\bibnamefont{Coleman}},
  \bibinfo{journal}{Nature} \textbf{\bibinfo{volume}{424}},
  \bibinfo{pages}{524} (\bibinfo{year}{2003}).

\bibitem[{\citenamefont{Feynman and Cohen}(1956)}]{PhysRev.102.1189}
\bibinfo{author}{\bibfnamefont{R.~P.} \bibnamefont{Feynman}} \bibnamefont{and}
  \bibinfo{author}{\bibfnamefont{M.}~\bibnamefont{Cohen}},
  \bibinfo{journal}{Phys. Rev.} \textbf{\bibinfo{volume}{102}},
  \bibinfo{pages}{1189} (\bibinfo{year}{1956}).

\bibitem[{\citenamefont{Zhang et~al.}(2011)\citenamefont{Zhang, Grover, and
  Vishwanath}}]{PhysRevLett.107.067202}
\bibinfo{author}{\bibfnamefont{Y.}~\bibnamefont{Zhang}},
  \bibinfo{author}{\bibfnamefont{T.}~\bibnamefont{Grover}}, \bibnamefont{and}
  \bibinfo{author}{\bibfnamefont{A.}~\bibnamefont{Vishwanath}},
  \bibinfo{journal}{Phys. Rev. Lett.} \textbf{\bibinfo{volume}{107}},
  \bibinfo{pages}{067202} (\bibinfo{year}{2011}).

\bibitem[{\citenamefont{Hastings et~al.}(2010)\citenamefont{Hastings, Gonzalez,
  Kallin, and Melko}}]{Hastings+10}
\bibinfo{author}{\bibfnamefont{M.~B.} \bibnamefont{Hastings}},
  \bibinfo{author}{\bibfnamefont{I.}~\bibnamefont{Gonzalez}},
  \bibinfo{author}{\bibfnamefont{A.~B.} \bibnamefont{Kallin}},
  \bibnamefont{and} \bibinfo{author}{\bibfnamefont{R.~G.} \bibnamefont{Melko}},
  \bibinfo{journal}{Phys. Rev. Lett.} \textbf{\bibinfo{volume}{104}},
  \bibinfo{pages}{157201} (\bibinfo{year}{2010}).

\bibitem[{\citenamefont{Calabrese et~al.}(2012)\citenamefont{Calabrese,
  Mintchev, and Vicari}}]{Calabrese+12}
\bibinfo{author}{\bibfnamefont{P.}~\bibnamefont{Calabrese}},
  \bibinfo{author}{\bibfnamefont{M.}~\bibnamefont{Mintchev}}, \bibnamefont{and}
  \bibinfo{author}{\bibfnamefont{E.}~\bibnamefont{Vicari}},
  \bibinfo{journal}{EPL} \textbf{\bibinfo{volume}{97}}, \bibinfo{pages}{20009}
  (\bibinfo{year}{2012}).

\bibitem[{\citenamefont{{Zaanen} et~al.}(2008)\citenamefont{{Zaanen},
  {Kr{\"u}ger}, {She}, {Sadri}, and {Mukhin}}}]{2008arXiv0802.2455Z}
\bibinfo{author}{\bibfnamefont{J.}~\bibnamefont{{Zaanen}}},
  \bibinfo{author}{\bibfnamefont{F.}~\bibnamefont{{Kr{\"u}ger}}},
  \bibinfo{author}{\bibfnamefont{J.~H.} \bibnamefont{{She}}},
  \bibinfo{author}{\bibfnamefont{D.}~\bibnamefont{{Sadri}}}, \bibnamefont{and}
  \bibinfo{author}{\bibfnamefont{S.~I.} \bibnamefont{{Mukhin}}},
  \bibinfo{journal}{Iranian Journal of Physics} \textbf{\bibinfo{volume}{8}},
  \bibinfo{pages}{39} (\bibinfo{year}{2008}), \eprint{arXiv:0802.2455}.

\bibitem[{\citenamefont{Swingle}(2010)}]{PhysRevLett.105.050502}
\bibinfo{author}{\bibfnamefont{B.}~\bibnamefont{Swingle}},
  \bibinfo{journal}{Phys. Rev. Lett.} \textbf{\bibinfo{volume}{105}},
  \bibinfo{pages}{050502} (\bibinfo{year}{2010}).

\bibitem[{\citenamefont{Swingle}(2012)}]{PhysRevB.86.035116}
\bibinfo{author}{\bibfnamefont{B.}~\bibnamefont{Swingle}},
  \bibinfo{journal}{Phys. Rev. B} \textbf{\bibinfo{volume}{86}},
  \bibinfo{pages}{035116} (\bibinfo{year}{2012}).

\bibitem[{\citenamefont{Paramekanti et~al.}(2002)\citenamefont{Paramekanti,
  Balents, and Fisher}}]{PhysRevB.66.054526}
\bibinfo{author}{\bibfnamefont{A.}~\bibnamefont{Paramekanti}},
  \bibinfo{author}{\bibfnamefont{L.}~\bibnamefont{Balents}}, \bibnamefont{and}
  \bibinfo{author}{\bibfnamefont{M.~P.~A.} \bibnamefont{Fisher}},
  \bibinfo{journal}{Phys. Rev. B} \textbf{\bibinfo{volume}{66}},
  \bibinfo{pages}{054526} (\bibinfo{year}{2002}).

\bibitem[{\citenamefont{Lai et~al.}(2013)\citenamefont{Lai, Yang, and
  Bonesteel}}]{PhysRevLett.111.210402}
\bibinfo{author}{\bibfnamefont{H.-H.} \bibnamefont{Lai}},
  \bibinfo{author}{\bibfnamefont{K.}~\bibnamefont{Yang}}, \bibnamefont{and}
  \bibinfo{author}{\bibfnamefont{N.~E.} \bibnamefont{Bonesteel}},
  \bibinfo{journal}{Phys. Rev. Lett.} \textbf{\bibinfo{volume}{111}},
  \bibinfo{pages}{210402} (\bibinfo{year}{2013}).

\bibitem[{\citenamefont{Lai and Yang}(2016)}]{PhysRevB.93.121109}
\bibinfo{author}{\bibfnamefont{H.-H.} \bibnamefont{Lai}} \bibnamefont{and}
  \bibinfo{author}{\bibfnamefont{K.}~\bibnamefont{Yang}},
  \bibinfo{journal}{Phys. Rev. B} \textbf{\bibinfo{volume}{93}},
  \bibinfo{pages}{121109} (\bibinfo{year}{2016}).

\end{thebibliography}
\end{document}